\documentclass[journal,twocolumn]{IEEEtran}

\usepackage{amsmath,amssymb,amsfonts,cuted}
\usepackage[final]{graphicx}
\usepackage{psfrag}
\usepackage{epsfig}
\usepackage[numbers,sort&compress]{natbib}

\usepackage{flushend}
\usepackage{array,booktabs}

\usepackage{amsmath,amssymb,amsfonts}
\usepackage[final]{graphicx}
\usepackage{flushend}
\usepackage[numbers,sort&compress]{natbib}

\usepackage{ucs} 
\usepackage[utf8x]{inputenc}
\usepackage[usenames,dvipsnames]{xcolor}
\usepackage{tikz}
\usepackage{tkz-tab}
\usepackage{pgfplots}
\usepackage{caption}
\usepackage{subcaption}

\usepackage{multirow}
\usepackage{latexsym}
\usepackage{amssymb}
\usepackage{amsmath}
\usetikzlibrary{positioning}

\begin{document}

\title{Wideband Ultrasonic Acoustic Underwater Channels: Measurements and Characterization}

\author{Jes\'us~L\'opez-Fern\'andez,~Unai Fern\'andez-Plazaola, Jos\'e~F.~Paris,~\IEEEmembership{Member,~IEEE}, Luis D\'iez, Eduardo~Martos-Naya}

\maketitle

\begin{abstract}
In this  work we present the results of a measurement campaign carried out in the Mediterranean sea aimed at characterizing the underwater acoustic channel in a wideband at ultrasonic frequencies centered at 80 kHz with a width of 96 kHz, covering two octaves from 32 to 128 kHz. So far, these type of wideband measurements are not found in the literature. Periodic orthogonal frequency division multiplexing (OFMD) sounding signals using Zadoff-Chu sequences have been specially designed for this purpose. The collected data has been post-processed to estimate the time-variant impulse and frequency responses and relevant parameters for system design like the time coherence, bandwidth coherence, delay spread and Doppler bandwidth. The statistical behavior of the channel gain random fluctuation has also been analyzed. This information has been extracted for both the global channel and each path separately. The wide bandwidth of the measurements have allowed the characterization of the channel in a scarcely explored ultrasonic band with an accuracy that is far beyond what is reported in previous works.
\end{abstract}

\begin{IEEEkeywords}
Underwater Acoustic Communications (UAC), underwater ultrasonic channels, shallow water acoustic channel, channel estimation, Zadoff-Chu.
\end{IEEEkeywords}

\section{Introduction}

The underwater Acoustic Communication (UAC) channel is one of the toughest wireless communication media in use today for several reasons, specially in the case of shallow waters. First, it presents a high attenuation which grows with frequency and suffers from hard multipath propagation due to multiple reflections on the seabed and water surface which turns into long delay impulse responses. Second, the low speed of sound in water gives rise to high Doppler broadening of the signals spectrum even with slow motion of the transceivers. The result is a channel which in many cases can be considered overspread \cite{Walree2010}, i.e., the product of the delay and Doppler spreads is greater than one. Third, a considerable variation of the channel characteristics may occur even with a slight change of location or weather conditions, which leads to the non-existence of a typical UAC channel \cite{StojanovicCommMag2009}.
 
The UAC channel has been traditionally exploited in the audio band (up to around 15 kHz) for low speed digital applications that operate using a bandwidth of only several kHz. However, in recent years real-time monitoring of marine environment and supervision of undersea equipment health of oil and gas companies using remotely or autonomous operated vehicles call for higher speed digital links, specially if high quality video transmission is required. Such higher speed digital systems demand bandwidths of several tens of kHz which are not available in the audio band. We must therefore move up in the acoustic spectrum (going into the ultrasonic band) where wider bandwidths can be exploited. Notice that moving up in frequency is a need but it has several drawbacks in the already complicated behavior of the underwater channel such as a higher attenuation (which will inevitably cause a shortening of the communication link range as compared to those operating at lower frequencies) and a reduction of both the coherence time and the coherence bandwidth. This poses a challenge in the design of systems that operate in this scenario. The first step for achieving this goal is the characterization of such channels through measurements which is the topic of this work.

The UAC channel has been extensively analyzed in the audio band (see \cite{StojanovicCommMag2009,Yang2006,Kim2013,Kulhandjian2014,Stojanovic2013,Kim2015} among many others) but less works explore the ultrasonic band. Specifically, measurements in the 2-32 kHz, 20-60 kHz, 37-41 kHz and 60-100 kHz bands are reported in \cite{Walree2013,Walree2013bis} \cite{Chitre2007}, \cite{Zhang2010} and \cite{Ochi2008}  respectively while measurements centered at a much higher frequency (820 kHz) are presented in \cite{Hajenko2010}. An important point worth mentioning is that in all these cited works, the measurements are narrowband (the bandwidth is between 1\% and 20\% of the center frequency), i.e. although the frequency range in which the measurements are carried out is broad, the bandwidth explored within that range is narrow. This is also the case of the results we presented in \cite{Canete2016} where we explored the fading behavior of the shallow water AUC channel in the range from 32 to 128 kHz by analyzing  the amplitude fluctuation of transmitted tones.

In this paper we present the results of wideband acoustic measurements at ultrasonic frequencies ranging from 32 to 128 kHz carried out in shallow waters of the Mediterranean Sea. The transmitter and receiver have been designed for that purpose as well as the sounding signals which consist in 96 kHz bandwidth periodic OFDM signals. The bandwidth-to-center frequency ratio of measurements is 120\%. A specific procedure for the estimation of the channel time-variant impulse response and time-variant frequency response has been developed. As a derivative result, functions like the time and frequency autocorrelation, the Doppler spectrum, the power delay profile and their derived parameters like the coherence time, coherence bandwidth, delay spread and Doppler spread as well as the characterization of the random fluctuation of the channel gain have been obtained corresponding to both the global channel and the individual separated paths (line-of-sight, surface reflected, bottom reflected, etc). Relevant discrepancies are appreciated when compared to the results reported in the literature for narrowband \cite{Stojanovic2013}. This distinctive behavior must be taken into account in the design of transceiver systems that make profit of the wideband and achieve high data rate transmission.

As a summary, the main contributions of this paper are:
\begin{enumerate}
\item Design and implementation of a specific purpose measurement equipment for operation in the 32-128 kHz band. 
\item Design of specific sounding signals consistent in periodic OFDM signals with a bandwidth of 96 kHz and a method for the time-variant channel response estimation based on filter banks instead of the habitual correlation methods. 

\item Wideband measurements of the UAC channel at ultrasonic frequencies centered at 80 kHz and a bandwidth of 96 kHz (two octaves) along with the corresponding characterization of the global channel and of individual separated paths. To the author's best knowledge, these type of measurements are not reported in the literature.
\end{enumerate}

The remainder of this paper is organized as follows. Section \ref{MeasurementScenario} presents the scenario and the equipment used to carry out the measurements. Section \ref{ChannelEstimationSystem} explains the sounding signals and the method to estimate the channel. In \ref{UACcharacterization} the ultrasonic wideband UAC is characterized, first globally and then each path separately. Finally, some conclusions are summarized in Section \ref{Conclusions}.
 
\section{Measurement scenario and equipment}
\label{MeasurementScenario}

\subsection{Measurement scenario} 

The experimental data used in this work were collected by the authors along a measurement campaign in 2017. Shallow waters of the Mediterranean Sea in La Algameca Chica (Cartagena, Spain) were chosen to perform ultrasonic wideband UAC measurements. The campaign extended along several sessions with common smooth sea conditions (World Meteorological Organization sea state Code 2, waves of less than 0.5 m). The vertical sound   speed profile was approximately constant with a sound speed of 1525 m/s.

\begin{figure*}[t!]
\hspace{-5mm}\includegraphics[width=\textwidth]{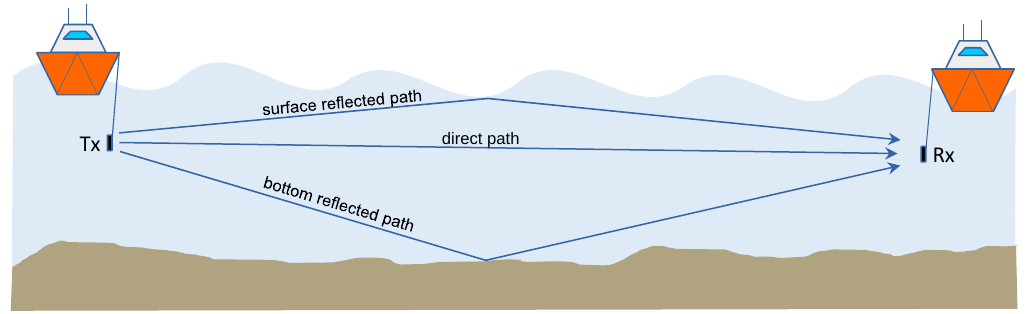}
	\caption{Measurement scenario.} \label{Fig_Escenario}
\end{figure*}

A total number of 13 channels have been measured in the 2 octave ultrasonic band from 32 kHz to 128 kHz. The projector (TX) and the hydrophone (RX) were submerged at a fixed depth of 6 m. from two different boats. Both boats were separated distances ranging from 50 to 400 m. approximately. The observed sea depth along the 13 channels spans from 19 to 34 m. The type of seabed in La Algameca Chica is sandy with some sparse rocks. A picture of the scenario is shown in Fig. \ref{Fig_Escenario} where the three main paths that the signal may follow are highlighted, i.e. the line-of-sight or direct path, the path corresponding to a reflection in the sea surface and the path that undergoes a reflection at the sea bottom.

\subsection{Measurement equipment}

\begin{figure}
\hspace{-5mm}\includegraphics[width=0.5\textwidth]{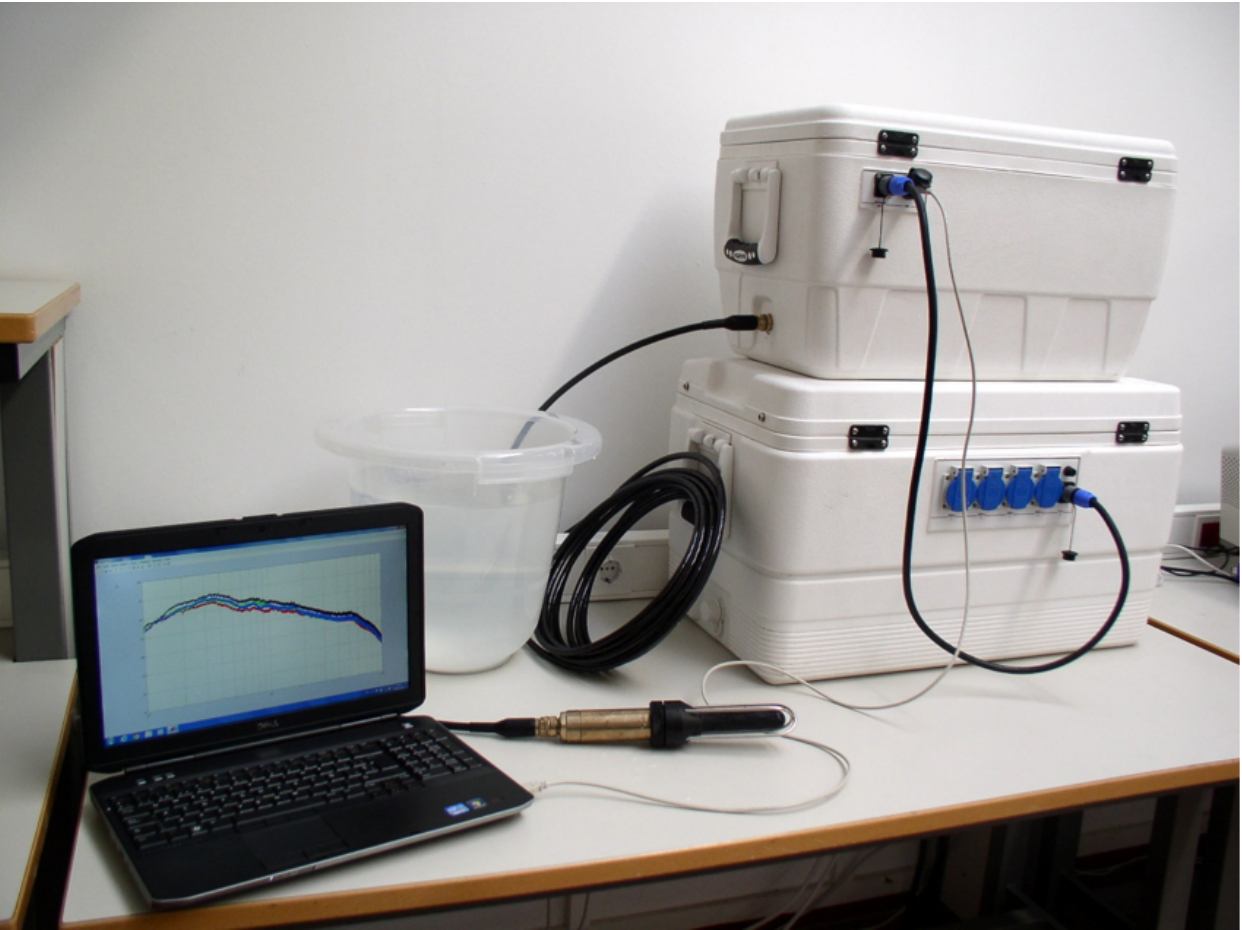}
	\caption{Measurement equipment at the receiver side.} \label{Fig_Equipos}
\end{figure}

\begin{figure}
\hspace{-5mm}\includegraphics[width=0.5\textwidth]{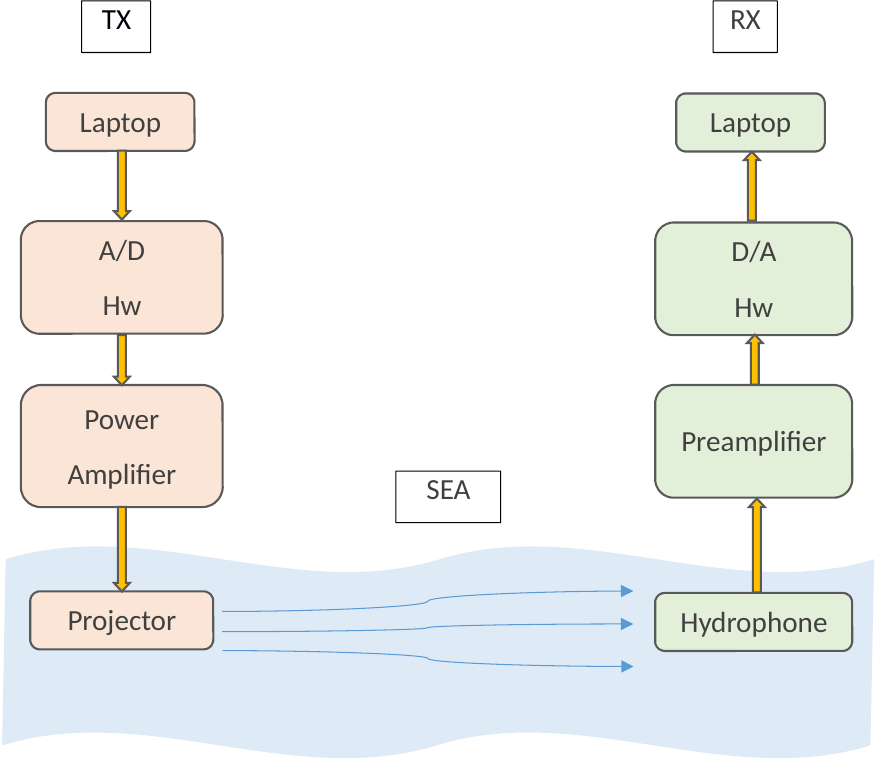}
	\caption{Block diagram of the measurement system.} \label{Fig_Esquema}
\end{figure}

A photograph of the measurement equipment is shown in Fig. \ref{Fig_Equipos} where the receiver side of the channel sounder is displayed in the laboratory of the University of Malaga (the transmitter side is similar). A functional scheme of the hardware used for the measurements is depicted in Fig. \ref{Fig_Esquema}. The equipment has been designed and assembled by the authors to investigate ultrasonic wideband UAC and was already used in the narrowband UAC measurement campaign reported in \cite{Canete2016}. 

The hardware includes: two laptops (for signal control, storage and off-line signal processing for monitoring purposes); two acquisition modules IOtech Personal DAQ3000 (with 16 bits of resolution and a 1-MHz sample rate, which were used as a digital-to-analog (D/A) converter at the transmitter and as an analog-to-digital (A/D) converter at the receiver); Brüel Kjaer 2713 power amplifier and Brüel Kjaer projector 8105 at the transmitter; Reson VP2000 voltage preamplifier EC6081 (with a 1-MHz bandwidth) and Reson TC4032 low noise hydrophone at the receiver. 

The above described hardware is operated by a software implemented by the authors which enables long measurements in an automatic way, i.e. complete training sequences of wideband signals for each channel are sent and the received signals are then recorded and automatically separated by the software. Afterward, the records are digitally post-processed using the algorithms described in section \ref{ChannelEstimationSystem}.


\section{Channel estimation system}
\label{ChannelEstimationSystem}
In this section the designed probe signals are described along with the corresponding signal processing method employed in the estimation of the channel response. First, the notation used in the forthcoming sections is detailed.  
\subsection{Notation preliminaries}
The average value of a signal $x(u)$ with respect to variable $u$ will be denoted as $<x(u)>_u$ and is given by
\begin{equation}
<x(u)>_u=\lim_{T\to \infty} \frac{1}{T}\int_{-T/2}^{T/2}x(u)du,
\end{equation}
in case $x(u)$ is a power signal and 
\begin{equation}
<x(u)>_u=\int_{-\infty}^{\infty}x(u)du,
\end{equation}
in case $x(u)$ is an energy signal. The Fourier transform of an energy signal $x(u)$ with respect to variable $u$ will be denoted as $\mathcal {FT}_u\{x(u)\}$ and can be expressed as
\begin{equation}
\mathcal {FT}_u\{x(u)\}=<x(u)e^{-j \psi u}>_u
\end{equation}
 where $\psi$ is the transformed domain variable. The inverse Fourier transform of $g(\psi)$ with respect to variable $\psi$ will be referred to as $\mathcal {FT}^{-1}_{\psi} \{g(\psi)\}$ and is defined accordingly
\begin{equation}
\mathcal {FT}^{-1}_{\psi} \{g(\psi)\}=<g(\psi)e^{j \psi u}>_{\psi}
\end{equation}
The autocorrelation of $x(u)$ with respect to $u$ will be written as $\Phi_x^{(u)}(\Delta u)$ and is defined as 
\begin{equation}
\Phi_x^{(u)}(\Delta u)=<x(u+\Delta u)x^*(u)>_u
\end{equation}
In the previous definitions the reference signal $x(u)$ can be not only a function of $u$ but a multivariate function where other variables are involved. This justifies the use of the subscript in the definition of the average, Fourier transform and autocorrelation as it indicates on which of the variables the operation is to be performed.

In addition, we denote the exact time-variant impulse and frequency responses of the channel as $h(t,\tau)$ and $H(t,f)$ respectively. The corresponding estimated responses are denoted as $\hat{h}(t,\tau)$ and $\hat{H}(t,f)$ while the responses obtained after initial delay compensation (explained below) are denoted as $\tilde{h}(t,\tau)$ and $\tilde{H}(t,f)$.

\subsection{Channel sounding signals}

For channel sounding we use periodic multitone signals which can be expressed as
\begin{equation}
	x(t)=\sum_{k=k_1}^{k_N} \cos(2\pi k \delta_{\!f} t +\Psi_k)=\sum_{k=k_1}^{k_N} \Re\{X_k e^{j(2\pi k\delta_{\!f}  t)}\} ,
	\label{Eq:DFS}
\end{equation}  
where  $\delta_{\!f}$ is the frequency separation between tones, $T=1/\delta_{\!f}$ is the fundamental period of $x(t)$, $X_k=e^{j\Psi_k}$, $\Psi_k$ are the relative phases of each tone, and the constants $k_1$ and $k_N$ delimit the channel sounding bandwidth.

For channel estimation purpose, the peak to average power ratio (PAPR) of the transmitted signal should be low to allow enough amplification while avoiding the signal getting into the non-linear range of the power amplifier. Taking into account that the expression of $x(t)$ in (\ref{Eq:DFS}) corresponds to a Fourier series with complex coefficients $X_k$, the PAPR can be minimized by designing those $X_k$ using a Zadoff-Chu sequence, as it is done in some digital communication standards for channel estimation \cite{Jurgen2013}. These sequences are defined as
\begin{equation}
 X_{k}={\text{exp}}\left(-j{\frac {\pi u k^2}{N_{\text{ZC}}}}\right),
\end{equation}
where $N_{ZC}$ is the length of the Zadoff-Chu sequence, and $u\in\mathbb{N}$ such that $\text{gcd}(u,N_{ZC})=1$ where $\text{gcd}(\cdot)$ stands for the greatest common divisor function. These sequences have constant envelope and, if $N_{ZC}$ is prime, their Discrete Fourier Transform (DFT) is also a Zadoff-Chu sequence and therefore, it has a constant envelope too. Although the continuous signal $x(t)$ obtained from a Zadoff-Chu sequence $X_k$ does not meet this property (since a Fourier series expansion is not equivalent to a DFT) the envelope of the resultant signal has a very low PAPR. 

\begin{figure}
  \centering
  \begin{subfigure}[b]{0.5\linewidth}
    \includegraphics[width=120pt]{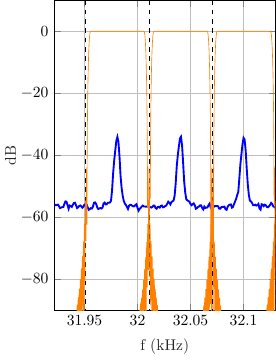}
		\vspace*{-2mm}
    \caption{\label{fig:espectro12}}
  \end{subfigure}%
  \begin{subfigure}[b]{0.5\linewidth}
    \includegraphics[width=120pt]{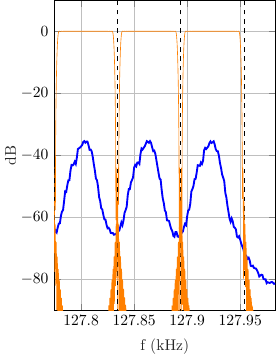}
		\vspace*{-2mm}
    \caption{\label{fig:espectro22}}
  \end{subfigure}
	 \caption{Partial spectrum of the received signal for the (a) first carriers and (b) last carriers. The frequency response of the filters $g_i[n]$ used for channel estimation is superimposed in orange color.}\label{SX1}	
  
\end{figure}

When the multitone signal $x(t)$ in (\ref{Eq:DFS}) is transmitted through a linear time-variant channel (LTV) characterized by its time-variant channel impulse response $h(t,\tau)$, the output can be expressed as 

\begin{equation}
\begin{split}
y(t)=& x(\tau)\ast h(t,\tau)=\int_{0}^{\infty}x(t-\tau)h(t,\tau)d\tau=\\
&\sum_{k=k_1}^{k_N}  \Re\{ e^{j(2\pi k \delta_{\!f} t+\Psi_k)}\int_{0}^{\infty}h(t,\tau) e^{-j2\pi k \delta_{\!f} \tau}d\tau\}=\\
		& \sum_{k=k_1}^{k_N} \Re\{H(t,k\delta_{\!f})e^{j(2\pi k \delta_{\!f} t+\Psi_k)}\},\\
\end{split}
\end{equation}
where $H(t,f)$ is the time-variant channel frequency response. The time autocorrelation of the time-variant frequency response can be defined as
\begin{equation}
\Phi_H^{(t)}(\Delta t,f)=<H(t+\Delta t,f)H^*(t,f)>_t
\end{equation}
and the spectral density of the time-variant frequency response is
\begin{equation}
R(\nu,f)=\mathcal{FT}_{\Delta t}\{ \Phi_H^{(t)}(\Delta t,f)\}.
\end{equation}
Notice that for a fixed frequency $f$ the function $R(\nu,f)$ is the Doppler spectrum of the channel frequency response at that frequency.
Assuming that $H(t,f)$ varies with time much slower than does a carrier of frequency $f$, it can be shown that the time autocorrelation of the received signal $y(t)$ can be approximated by (see Appendix)
\begin{equation} \label{Eq:Autocorr_Salida}
\Phi_y^{(t)}(\Delta t) \approx \tfrac{1}{2}\sum_{k=k_1}^{k_N}  \Re\{\Phi_H^{t}(\Delta t, k \delta_{\!f})e^{j2\pi k \delta_{\!f} \Delta t }\}
\end{equation}
and the power spectral density of the received signal can be expressed as
\begin{equation}
P_y(f) \approx \sum_{k=k_1}^{k_N} \tfrac{1}{4}R(f-k\delta_{\!f},k\delta_{\!f})+\tfrac{1}{4}R^*(-f-k\delta_{\!f},k\delta_{\!f}).
\end{equation}
One of the key parameters in the design of the sounding signals is $\delta_{\!f}$. If $\delta_{\!f}$ is too large, the frequency sampling period of the channel response estimation could be insufficient and could lead to aliasing in the estimated time domain response. Conversely, if $\delta_{\!f}$ is too small, the spectrum broadening of the received tones due to Doppler effect could be large compared to $\delta_{\!f}$, inducing overlap between the spectrum of adjacent tones yielding a degradation of the channel response estimation. Moreover, since the total power of the sounding signal is distributed among the tones, a small $\delta_{\!f}$ (this is, a high number of tones), will make the signal to noise ratio (SNR) of each tone component lower, degrading channel estimation. In Fig. \ref{SX1} it can be observed the spectrum $P_y(f)$ of the received signal, where it can be appreciated that the Doppler spread of the first carriers (Fig. \ref{fig:espectro12}) is considerably narrower than that of the last carriers (Fig. \ref{fig:espectro22}).

Since the duration of the channel impulse response is a priori unknown, a set of sounding signals with different parameters has been employed in an initial measure and the one which has given the best performance has been selected for the measurement campaign presented in this work. The final selected parameters have been: $\delta_{\!f}$ = 333.3 Hz, $N_{ZC}=97$, $u=3$, $k_0=98$ and $k_N=194$ covering the band between 32 kHz and 128 kHz. The sounding signal period is 16.8 ms For every location, the signal is sent for 60 s. which results in 3571 consecutive channel estimations.

\subsection{Channel estimation method}
\label{Channel estimation method}
\begin{figure*}[t!]
\centering
\includegraphics[scale=1.2]{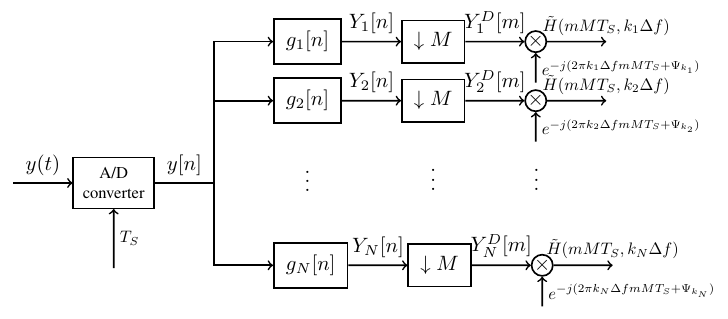} 
		\caption{Block diagram of the signal processing system for the channel frequency response estimation based on a bank of filters.} \label{SPS}
\end{figure*}

At the receiver, the signal is sampled with a period $T_S$ and the time-variant frequency response of the channel is obtained using a bank of $N$ filters centered at the frequency $k_i\delta_{\!f}$, $i=1...N$. A block diagram of the procedure is depicted in Fig. \ref{SPS}. The complex impulse response of each filter can be expressed as 
\begin{equation}
 g_i[n]=g[n]e^{j2\pi k_i \delta_{\!f} n},
\end{equation}
where $g[n]$ is the impulse response of a discrete-time finite impulse response (FIR) low-pass filter with cut-off frequency $\delta_{\!f}/2$. The digital filter bank and the decimation are efficiently implemented by windowing the received signal (taking $g[n]$ as window) and using Fast Fourier Transformation (FFT) \cite{Vaidyanathan1993}.

Assuming negligible noise level, low tone spectrum overlap and close-to-ideal frequency responses of the filters, the complex output signal of each filter can be approximated by
\begin{equation}
 Y_i[n] \approx H(nT_s,k_i\delta_{\!f})e^{j(2\pi k_i\delta_{\!f} n T_S+\Psi_k)}
\end{equation}
and its decimated version by
\begin{equation}
Y_i^D[m]  \approx H(m M T_s,k_i\delta_{\!f})e^{j(2\pi k_i\delta_{\!f} m M T_S+\Psi_k)}.
\label{Eq:DecimatedY}
\end{equation}

With regard to the decimation factor $M$, its value must be set so that aliasing in the output signal is avoided. Concretely, $M$ will be determined by the bandwidth of the filters (which is approximately $\delta_{\!f}$) and the sampling period $T_S$. Applying the Nyquist criterion yields \mbox{$M<\frac{1}{\delta_{\!f} T_S}$}. The estimated channel frequency response $\hat{H}(t,f)$ is obtained by demodulating the decimated signal in (\ref{Eq:DecimatedY})
\begin{equation}
\label{Eq:Estimacion_H}
	\hat{H}(mMT_s,k_i\delta_{\!f})=Y_i^D[m]e^{-j(2\pi k_i\delta_{\!f} m M T_S+\Psi_{k_i})}.
\end{equation}

Finally, the non-flat frequency response of the projector and hydrophone are compensated by using the frequency response curve provided by the manufacturer. From $\hat{H}(t,f)$, the estimated time-variant impulse response of the channel $\hat{h}(t,\tau)$ is obtained using inverse discrete Fourier transformation.

\subsection{Initial delay compensation}

The channel estimation described in the preceding sections is subject to  Doppler effects caused by two factors. On one side the frequency offset between the sampling clocks at the A/D and D/A converters at the transmitter and receiver respectively which results in a linear phase drift (with the corresponding Doppler shift) even with very small (few parts per million) offsets. On the other side is the unintentional transmitter/receiver motion, water surface motion and in general terms, the variation of the propagation medium itself which give rise to a (randomly) time varying phase shift leading to Doppler spreading. The combined phase shifts result in a fluctuation of the initial delay of the estimated impulse responses of the channel that must be corrected before any other processing \cite{Walree2013}. The procedure to achieve this goal is carried out as follows: first an initial frequency channel response estimation $\hat{H}(t,f)$ is performed as indicated in (\ref{Eq:Estimacion_H}). By means of the inverse discrete Fourier transform, the impulse response of channel $\hat{h}(t,\tau)$ is obtained and from it, the initial delay variation and the corresponding time-variant sampling frequency offset are estimated. Zero-padding in frequency domain is used in order to improve the precision in the delay estimation. Once the sampling frequency offset is available, the received signal is resampled and new improved channel impulse response $\tilde{h}(t,\tau)$ is obtained, i.e. $\tilde{h}(t,\tau)=\hat{h}(t,\tau-\tau_0(t))$ where $\tau_0(t)$ is the estimated initial delay variation.

\begin{figure*}
    \centering
    \begin{subfigure}[b]{0.5\textwidth}
        \includegraphics[width=\textwidth]{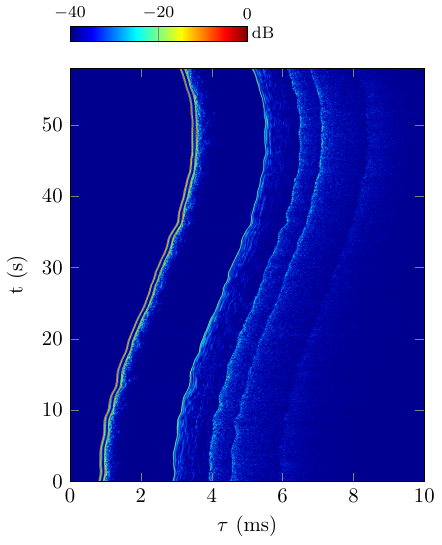}
        \caption{}
        \label{fig:IR2Da}
    \end{subfigure}%
		\centering
    \begin{subfigure}[b]{0.5\textwidth}
        \includegraphics[width=\textwidth]{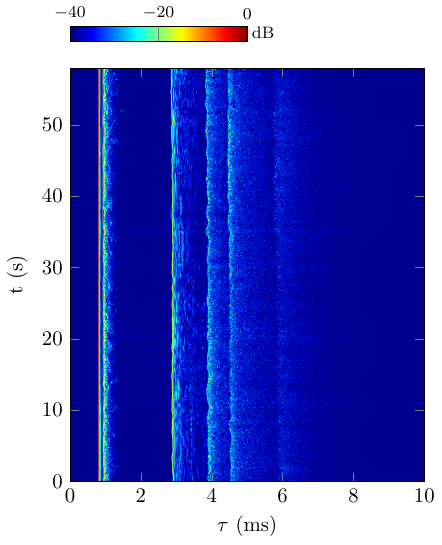}
        \caption{}
        \label{fig:IR2Db}
    \end{subfigure}
		\caption{Estimated time-variant impulse response of one of the measured channels (a) before initial delay compensation $|\hat{h}(t,\tau)|$ and (b) after initial delay compensation $|\tilde{h}(t,\tau)|$.} \label{TR1}
\end{figure*}

\begin{figure}
  \centering
  \begin{subfigure}[b]{0.5\linewidth}
    \centering\includegraphics[width=120pt]{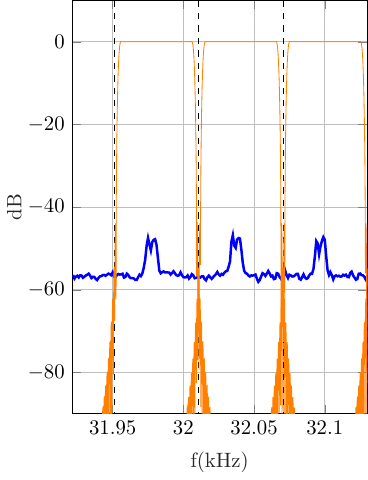}
		\vspace*{-2mm}
    \caption{\label{fig:espectro14}}
  \end{subfigure}%
  \begin{subfigure}[b]{0.5\linewidth}
    \centering\includegraphics[width=120pt]{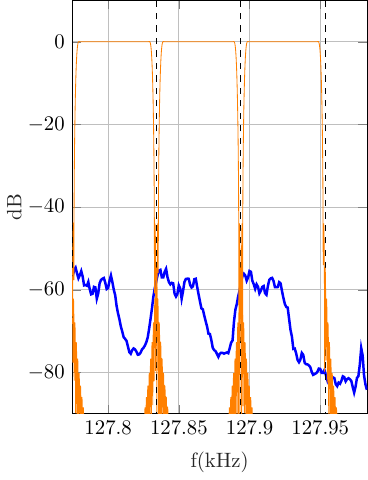} 
		\vspace*{-2mm}
    \caption{\label{fig:espectro24}}
  \end{subfigure} \\
	\vspace{4mm}
	 \begin{subfigure}[b]{0.5\linewidth}
    \centering\includegraphics[width=120pt]{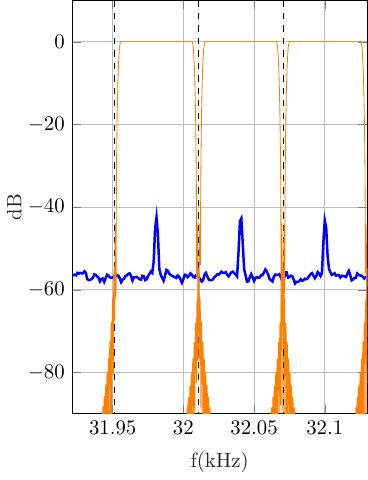}
		\vspace*{-2mm}
    \caption{\label{fig:espectro34}}
  \end{subfigure}%
  \begin{subfigure}[b]{0.5\linewidth}
    \centering\includegraphics[width=120pt]{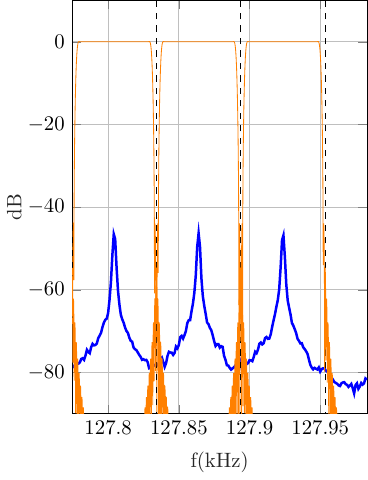} \\
		\vspace*{-2mm}
    \caption{\label{fig:espectro44}}
  \end{subfigure}
		
	 \caption{Partial spectrum (first and last carriers) of a received signal.  (a), (b) before and (c), (d) after initial delay compensation. In orange color, it is shown the frequency response of filter bank used for channel estimation.}\label{SX2}	
  
\end{figure}

In Fig. \ref{TR1} we can observe the time-variant channel impulse response $\hat{h}(t,\tau)$, from which the initial delay is estimated, and the time-variant channel impulse response $\tilde{h}(t,\tau)$ obtained after initial delay compensation of the received signal. Figures \ref{fig:espectro14} and \ref{fig:espectro24} show the signal spectra before initial delay compensation $\hat{H}(t,f)$ while in Fig. \ref{fig:espectro34} and \ref{fig:espectro44} the corresponding signal spectra after initial delay compensation $\tilde{H}(t,f)$ are depicted. It is evident the remarkable narrowing of the Doppler spread of each tone.

\section{Characterization of the wideband UAC}
\label{UACcharacterization}
\label{Characterization of the wideband UAC}

For convenience, we use continuous-time notation although $\tilde{H}(t,f)$ -and $\tilde{h}(t,\tau)$- are only known in a finite number of separated instants and frequencies as indicated in section \ref{Channel estimation method}. 

The measured channels are analyzed attending to various functions and parameters related to $\tilde{h}(t,\tau)$ and $\tilde{H}(t,f)$, i.e. the delay spectrum, the Doppler spectrum and the time and frequency autocorrelations. From these functions relevant parameters for systems design like delay spread, coherence bandwidth, Doppler spread and coherence time can be extracted. 

The time autocorrelation denoted as $\Phi^{(t)}(\Delta t)$ is defined as 
\begin{equation}
\label{Eq:Autocorr_t_1}
\Phi^{(t)}(\Delta t)=<\Phi_{\tilde{h}}^{(t)}(\Delta t, \tau)>_\tau ,
\end{equation}
where $\Phi_{\tilde{h}}^{(t)}(\Delta t, \tau)$ is the time autocorrelation of the time-variant channel impulse response which is defined as
\begin{equation}
\label{Eq:Autocorr_t_2}
\Phi_{\tilde{h}}^{(t)}(\Delta t, \tau)=<\tilde{h}(t+\Delta t,\tau)\tilde{h}^*(t,\tau)>_t ,
\end{equation}
so that $\Phi^{(t)}(\Delta t)$ corresponds to the average of the different time autocorrelations $\Phi_{\tilde{h}}^{(t)}(\Delta t, \tau)$ obtained for each $\tau$. 
Accordingly, the frequency autocorrelation denoted as $\Phi^{(f)}(\Delta f)$ is given by 
\begin{equation}
\label{Eq:Autocorr_f_1}
\Phi^{(f)}(\Delta f)=<\Phi_{\tilde{H}}^{(f)}(t, \Delta f)>_t ,
\end{equation}
where $\Phi_{\tilde{H}}^{(f)}(t, \Delta f)$ is the frequency autocorrelation of the time-variant channel frequency response which is defined as
\begin{equation}
\label{Eq:Autocorr_f_2}
\Phi_{\tilde{H}}^{(f)}(t, \Delta f)=<\tilde{H}(t, f+\Delta f)\tilde{H}^*(t,f)>_f.
\end{equation}
In this case $\Phi^{(f)}(\Delta f)$ corresponds to the average of the different frequency autocorrelations $\Phi_{\tilde{H}}^{(f)}(t, \Delta f)$ obtained for each $t$. 

 The Doppler spectrum $P_{Doppler}(\nu)$ and the power delay spectrum $P_{delay}(\tau)$ are Fourier-transform pairs of the time and frequency autocorrelations $\Phi^{(t)}(\Delta t)$ and $\Phi^{(f)}(\Delta f)$, concretely
\begin{align}
&P_{Doppler}(\nu)=\mathcal{FT}\{\Phi^{(t)}(\Delta t)\} \label{Eq:TF_dop},   \\
&P_{delay}(\tau)=\mathcal{FT}^{-1}\{\Phi^{(f)}(\Delta f)\} \label{Eq:TF_dly}.    
\end{align}

An alternative way of computing $P_{delay}(\tau)$ and $P_{Doppler}(\nu)$ is through the well-known scattering function \cite{Bello1963} $S(\nu,\tau)=\left|\mathcal{FT}_t\{\tilde{h}(t,\tau)\} \right|^2$ that indicates the Doppler spectrum for each delay $\tau$. Averaging $S(\nu,\tau)$ with respect to the Doppler frequency $\nu$ and the delay $\tau$ we get 
\begin{align}
&P_{Doppler}(\nu)=<S(\nu,\tau)>_\tau \label{Eq:P_DOP},\\
&P_{delay}(\tau)=<S(\nu,\tau)>_\nu \label{Eq:P_DLY}.
\end{align}

Finally, using Parseval's relation  it can be shown that $P_{delay}(\tau)$ in (\ref{Eq:P_DLY}) can also be obtained in a more straightforward way as $P_{delay}(\tau)= <\left| \tilde{h}(t,\tau) \right|^2>_t$.

As stated earlier, we can extract relevant channel parameters from the functions defined above. In particular, the root mean squared (rms) Doppler spread $\sigma_{\nu}$ and rms delay spread $\sigma_{\tau}$ are defined as the rms width of $P_{Doppler}(\nu)$ and $P_{delay}(\tau)$ respectively and give a measure of the width of the corresponding function. Given that the rms width is the standard deviation of a function taken as a probability density function (pdf), we can write for $\sigma_{\tau}$
\begin{equation}
\sigma_{\tau}=\left[\int_{-\infty}^{\infty}\tau^2 g(\tau)d\tau-\left( \int_{-\infty}^{\infty}\tau g(\tau)d\tau \right)^2 \right]^{\frac{1}{2}}
\label{Eq:Delay spread}
\end{equation}
with $g(\tau)=P_{delay}(\tau)/\int_{-\infty}^{\infty} P_{delay}(\tau) d\tau$. The parameter $\sigma_{\nu}$ is analogously defined. 

A high Doppler spread $\sigma_{\nu}$ implies a channel with fast temporal fluctuation while a high delay spread $\sigma_{\tau}$ implies a channel with strong frequency selectivity. Another two relevant parameters are the coherence time $t_c$ and coherence bandwidth $b_c$ that also measure the  temporal fluctuation and the  frequency selectivity of a channel respectively and can be extracted from the width of the autocorrelations, $\Phi^{(t)}(\Delta t)$ and $\Phi^{(f)}(\Delta f)$ respectively, as the value where the autocorrelation falls below a certain threshold. In general terms an inverse relation exists between Doppler spread and coherence time on one side and between delay spread and coherence bandwidth on the other.

A remark applies before going any further. The functions defined above are deterministic in the sense that no statistical expected values are involved. Each of them has its ensemble average counterpart. In our study we won't embark in the formal statistical analysis of wide-sense stationarity uncorrelated scattering channels \cite{Bello1963}. In fact, the low number of realizations of each channel impulse response obtained with our measurements would not allow a reliable verdict. We therefore focus our attention in showing the propagation characteristics using a deterministic approach, not in applying the WSSUS formalism as such. 

In the following two sections we present the analysis of the measurements corresponding to two different scales. In the first analysis scale we treat the channel response as a whole while in the second scale we analyze each of the multipath components of the total response separately. In the first scale we will focus our attention on the time variation of the channel through the Doppler spread/coherence time parameter. The delay spread/coherence bandwidth are presumed to give less information as they will be highly determined by the geometry of the scenario in which the strongest multipath components are the direct and the surface reflected paths. In the second scale the measurements of the delay spread/coherence bandwidth of each multipath component may reveal important information that has not been reported in the literature. Moreover, results of the statistical distribution of the channel gain fluctuation are also presented in both scales which is a valuable information about the behavior of the underwater channel in the short time scale.

The nominal geometry of the 13 sounded channels whose results are presented in this work is shown in table \ref{Tab:Channels}. The channels are numbered in ascending order of separation between transmitter and receiver. The depth of both transmitter and receiver was fixed at 6 meters in all the measurements. The coarse measurement of the separation between the boats carried out on the sea has been later fine tuned from the recorded data by using the time difference between the first and second paths. This is the reason for the irregular grid of distances.     

\begin{table}     
\caption{Channels parameters (tx. and rx. depth = 6m).}             
\centering                     
\begin{tabular}{|c|c|c|}       
\hline
Channel & Separation (m)  & Sea depth (m)\\ \hline \hline                         
1 & 47 & 24 \\  
\hline                         
2 & 51 & 19 \\  
\hline                         
3 & 98 & 20 \\  
\hline                         
4 & 100 & 20 \\ 
\hline                         
5 & 134 & 22 \\ 
\hline                         
6 & 168 & 24 \\ 
\hline                         
7 & 197 & 20 \\ 
\hline                         
8 & 216 & 28 \\ 
\hline                         
9 & 236 & 29 \\ 
\hline                         
10 & 242 & 34 \\
\hline                         
11 & 248 & 25 \\
\hline                         
12 & 260 & 28 \\
\hline                         
13 & 387 & 31 \\
\hline                         
\end{tabular}                  
\label{Tab:Channels}  
\end{table}

\subsection{Global channel characterization }
\label{Global channel characterization}

\begin{figure}
\includegraphics[width=0.5\textwidth]{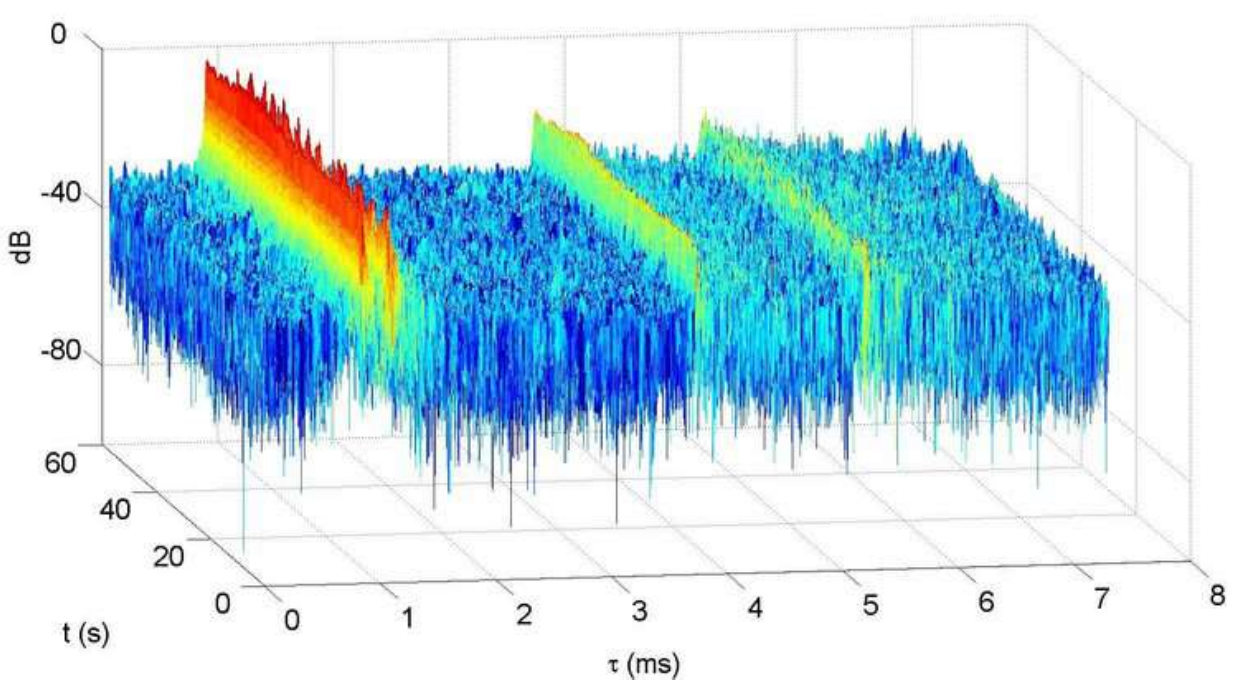}
\caption{Absolute value of the impulse response $|\tilde{h}(t,\tau)|$ as a function of $\tau$ and $t$ measured in channel 8 in logarithmic scale (the amplitudes are normalized to their maximum value).} 
\label{Fig_Canal_8_global_t}
\end{figure}

To start with, a complete measure is shown in Fig. \ref{Fig_Canal_8_global_t} where the square amplitude of the time-variant impulse response $|\tilde{h}(t,\tau)|^2$ is depicted for channel 8 over a duration of 60 s where the first four paths (corresponding to the direct, sea-reflected, bottom-reflected and sea-bottom-reflected paths) are clearly distinguishable.

\begin{figure}
  \centering
  \begin{subfigure}[b]{1\linewidth}
    \includegraphics[width=0.95\textwidth]{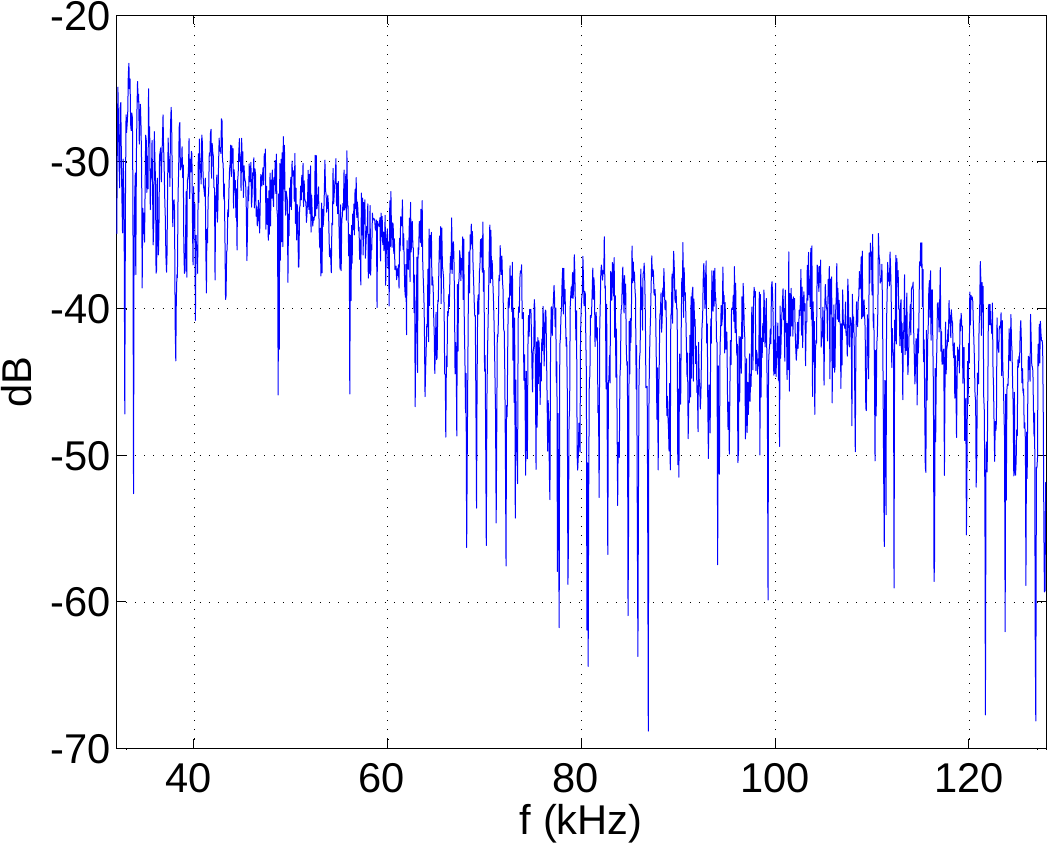}
    \caption{\label{fig:Hcompleta_canal_1}}
  \end{subfigure} \\
	\vspace{4mm}
  \begin{subfigure}[b]{1\linewidth}
    \includegraphics[width=0.95\textwidth]{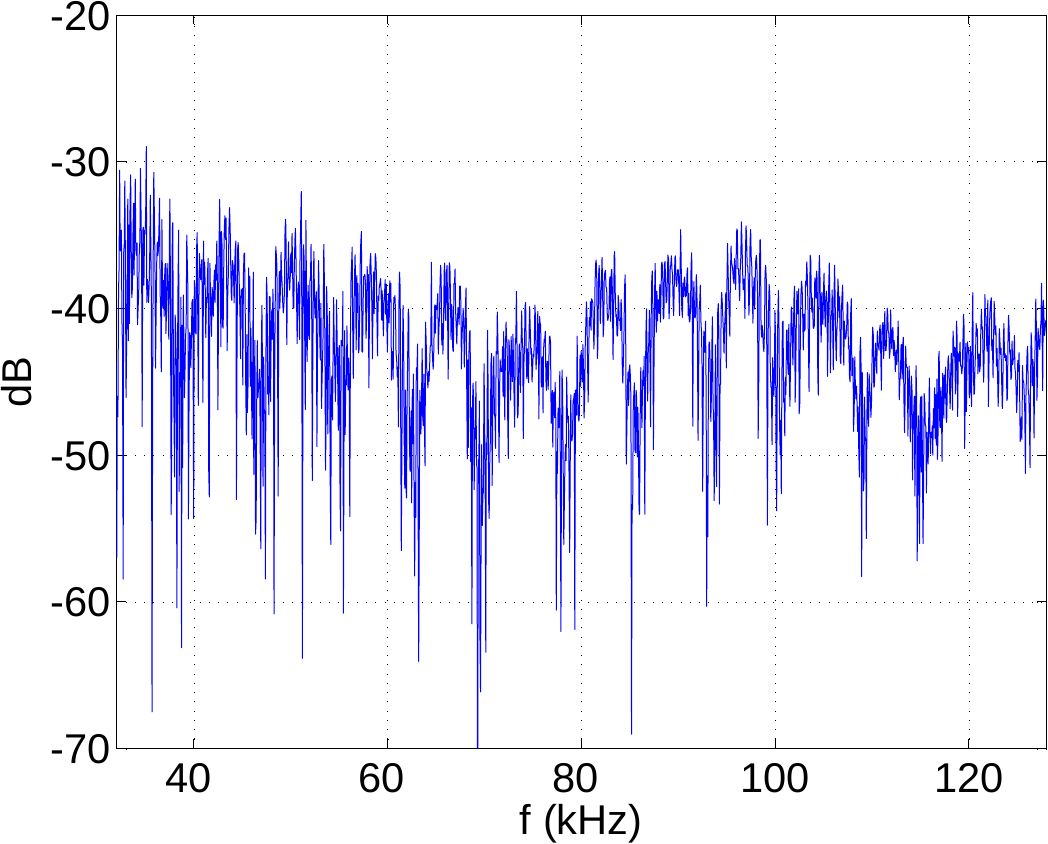}
    \caption{\label{fig:Hcompleta_canal_13}}
  \end{subfigure}
	 \caption{Absolute value of the frequency response $\tilde{H}(t,f)$ of (a) channel 1 and (b) channel 13 at a fixed $t$.}
	\label{Fig_Completa_2x2_H}
\end{figure}

In Fig. \ref{fig:Hcompleta_canal_1} and \ref{fig:Hcompleta_canal_13} the absolute value of the measured frequency response $|\tilde{H}(t,f)|$ corresponding to a fixed time $t$ in channel 1 and 13 respectively is depicted in the frequency band under study from 32 to 128 kHz. The attenuated comb-filter shape response is typical of a system dominated by two main paths, as it is the case. See that the lobe separation in channel 13 is greater than in channel 1 because so is the distance between transmitter and receiver and therefore, the time lapse between the first two paths is shorter.

\begin{figure*}
\centering
\includegraphics[width=\textwidth]{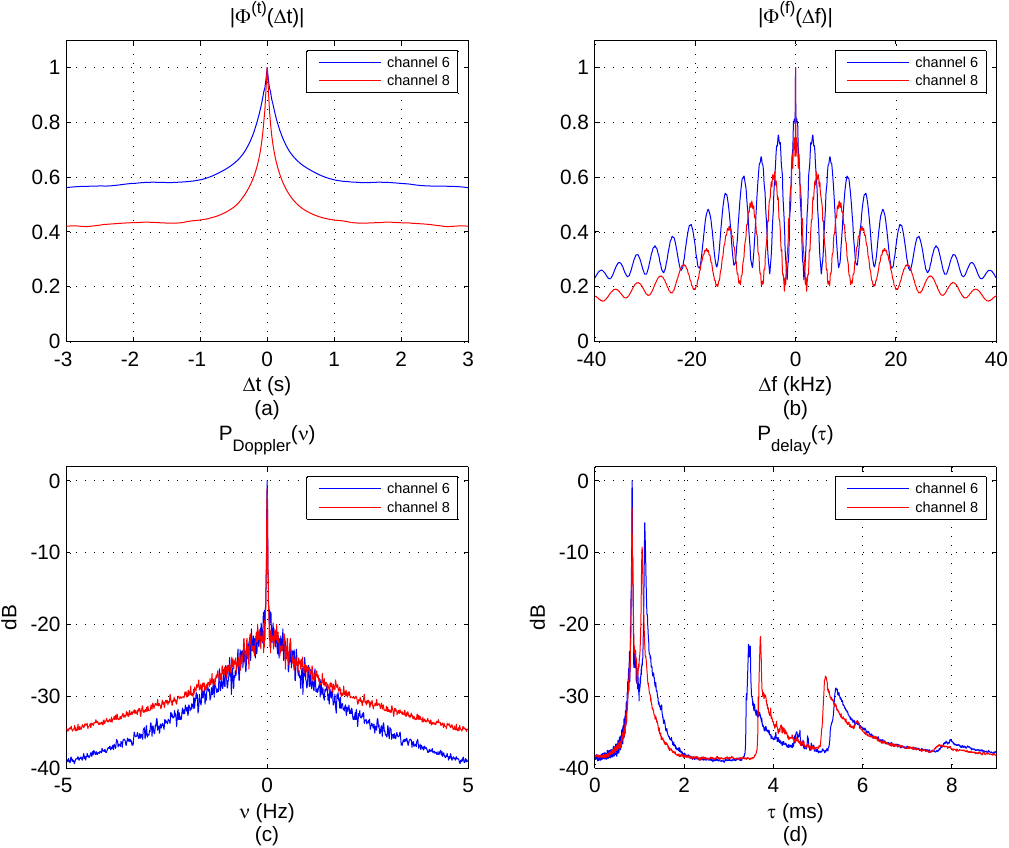}
\centering
\caption{(a) Absolute value of the time autocorrelation $|\Phi^{(t)}(\Delta t)|$, (b) frequency autocorrelation $|\Phi^{(f)}(\Delta f)|$, (c) Doppler spectrum $P_{Doppler}(\nu)$ and (d) normalized power delay profile $P_{delay}(\tau)$ of channels 6 and 8.} 
\label{Fig_Completa_2x2_4funciones_canal6}
\end{figure*}

In Fig. \ref{Fig_Completa_2x2_4funciones_canal6} the absolute value of the time autocorrelation $|\Phi^{(t)}(\Delta t)|$ in (\ref{Eq:Autocorr_t_1}), the frequency autocorrelation $|\Phi^{(f)}(\Delta f)|$ in (\ref{Eq:Autocorr_f_1}), the Doppler spectrum $P_{Doppler}(\nu)$ in (\ref{Eq:TF_dop}) and the delay spectrum $P_{delay}(\tau)$ in (\ref{Eq:TF_dly}) are depicted for channels 6 and 8. Starting with channel 6, see that $|\Phi^{(t)}(\Delta t)|$ falls off rapidly from 1 to 0.6 and remains practically constant thereafter which traduces in a Doppler spectrum $P_{Doppler}(\nu)$ with a sharp peak at 0 Hz reflecting the static part of the channel and a fast decaying behavior in the range of $\pm$ 5 Hz. The frequency autocorrelation $|\Phi^{(f)}(\Delta f)|$ features a decaying oscillation with a period inversely related of the time arrival difference between the two strongest paths, as expected. This is corroborated by the delay spectrum $P_{delay}(\tau)$ where the two first paths (direct and surface reflected) have a remarkably higher amplitude than the rest of the paths. A close look to the central lobe of $|\Phi^{(f)}(\Delta f)|$ reveals a sharp peak caused by the noise floor in the delay spectrum.

Regarding channel 8, the absolute value of the time autocorrelation $|\Phi^{(t)}(\Delta t)|$ falls off more rapidly than in channel 6 (what means a smaller coherence time) and stabilizes at a lower value of 0.4 what turns into a wider Doppler spectrum. The frequency autocorrelation $|\Phi^{(f)}(\Delta f)|$ shows a similar behavior but with a longer period of oscillation because the delay between the first two paths is shorter, as seen in the delay spectrum.

Table \ref{table:Global channel parameters} summarizes the coherence time $t_c$ and bandwidth $b_c$ measured for all the channels. The stabilization of $|\Phi^{(t)}(\Delta t)|$ in Fig. \ref{Fig_Completa_2x2_4funciones_canal6} in a constant high value is a general behavior, so we have used a high threshold of 0.9 in order to determine $t_c$. On the other hand, the sharp peak of $|\Phi^{(f)}(\Delta f)|$ at $\Delta f$ = 0 hinders the measure of $b_c$ since the main lobe is remarkably distorted. We have therefore adopted an alternative criteria for determining $b_c$ consistent in assigning to it the value of the frequency location of the first sidelobe peak. The delay spread of each channel computed following (\ref{Eq:Delay spread}) is also shown in Table \ref{table:Global channel parameters}. See that there is no evidence of an inverse relation between $b_c$ and $\sigma_{\tau}$ as it occurs in some type of terrestrial electromagnetic channels. The presence of several strong dispersive paths along with the diffuse components may be a reason for this.

\begin{figure}
  \centering
  \begin{subfigure}[b]{1\linewidth}
    \includegraphics[width=1\textwidth]{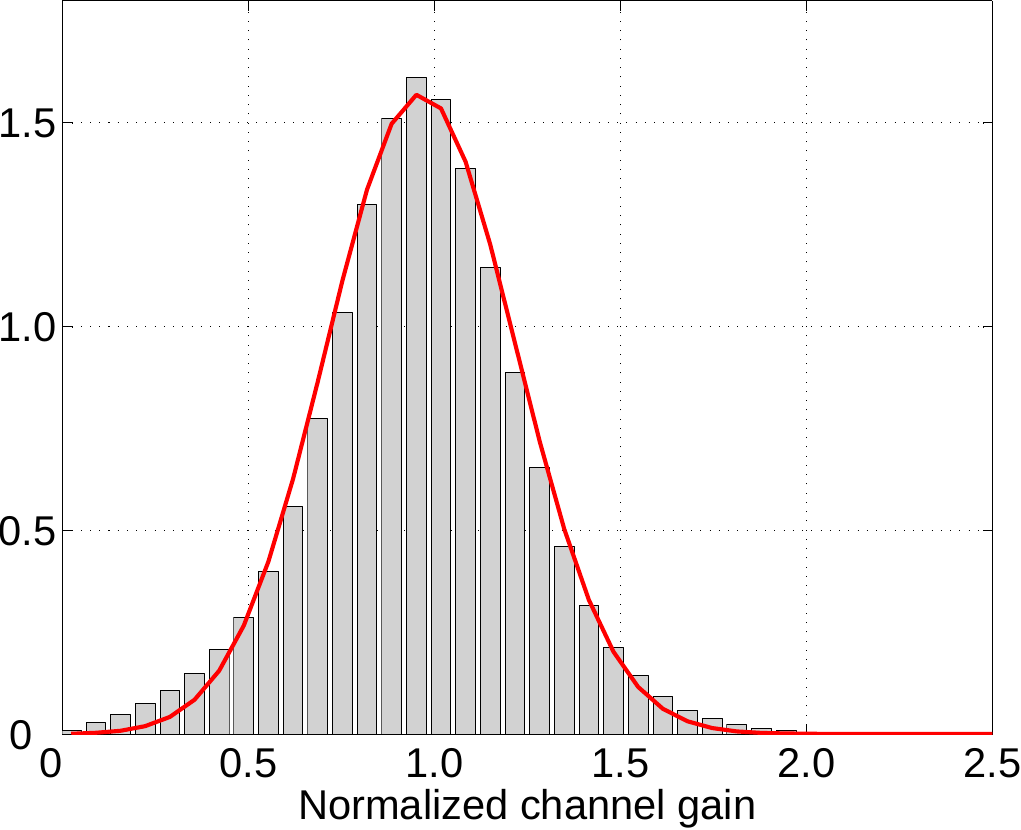}
    \caption{\label{fig:Histo_canal_2}}
  \end{subfigure} \\
	\vspace{4mm}
  \begin{subfigure}[b]{1\linewidth}
    \includegraphics[width=1\textwidth]{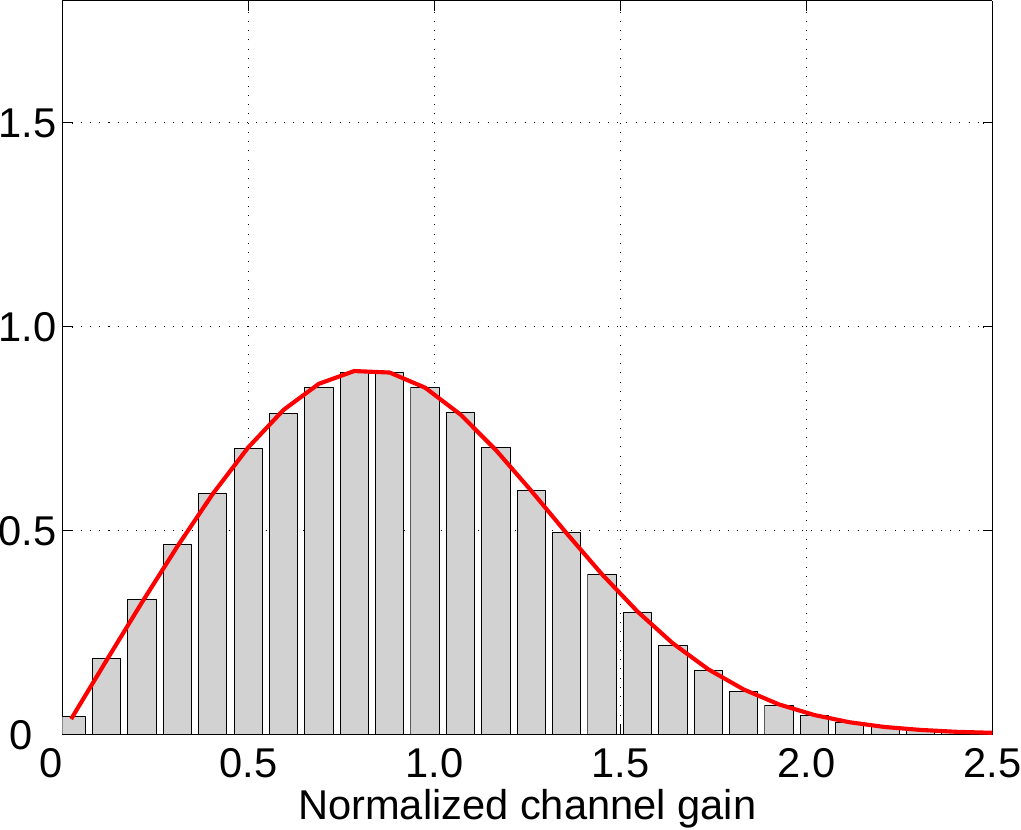}
    \caption{\label{fig:Histo_canal_3}}
  \end{subfigure}
		 \caption{Histogram of normalized channel gains and fitted Rician pdf for (a) channel 2 and (b) channel 3.}
	\label{Fig_Completa_2x2_HISTOGRAMA}
\end{figure}

The short term fluctuation of the channel gain has been analyzed by computing the histogram of $|\tilde{H}(t,f)|$ in short segments of time $t$ in order to average out the slow time variations. Additionally, narrow spectral segments have been selected for each time segment so that the channel gain fluctuation with frequency (due to frequency selectivity) is also averaged out. The histograms of the normalized gains for channels 2 and 3 are depicted in Fig. \ref{Fig_Completa_2x2_HISTOGRAMA}. A fit to a Rician distribution \cite{simon2005} has been carried out using the mean square error (MSE) criterion. Since the gains are normalized, the Rician distribution is dependent only on the Rician-K parameter. Denoting $x$ as the normalized gain, $p_{Rice}(x;K)$ as the Rician pdf and $p(x)$ as the histogram of the measured gains respectively, the fit is conveyed by finding the value of K that minimizes the MSE as \footnote{In this particular problem the MSE criterion has led to a better fit than that obtained other well-known methods like the K-S Test \cite{Cirrone2004} or the Jensen-Shannon divergence \cite{Lin1991}.}
\begin{equation}
\underset{K}{\text {min}}<\left|p_{Rice}(x;K)-p(x)\right|^2>_x
\label{Eq:MSE}
\end{equation}
The root of the minimum MSE (MMSE) achieved divided by the mean square value of $p(x)$ is used as relative fit accuracy indicator $\epsilon$, i.e. $\epsilon=\sqrt{\text{MMSE}/<|p(x)|^2>_x}$. The resulting Rician curves are superimposed in red color in Fig. \ref{Fig_Completa_2x2_HISTOGRAMA}. Notice the disparity of statistical behavior between the two selected channels. This is quantitatively measured in the last two columns of table \ref{table:Global channel parameters} where the fitted K factor and the relative fit error $\epsilon$ obtained in the fit for all the channels are shown. A high K factor indicates a stable arrival and therefore a predominance of the direct path over the rest and vice versa. In the measured channels the K factor ranges from values lower than 1 (see channels 3 and 7) that correspond to a close-to-Rayleigh distribution to values over 6 (channel 2). In general terms the K factor is low what means that the inherent dispersion associated to the reflected paths dominate the statistical behavior.

\begin{figure}[t]
  \centering
  \begin{subfigure}[t]{1\linewidth}
    \includegraphics[width=1\textwidth,height=0.81\textwidth]{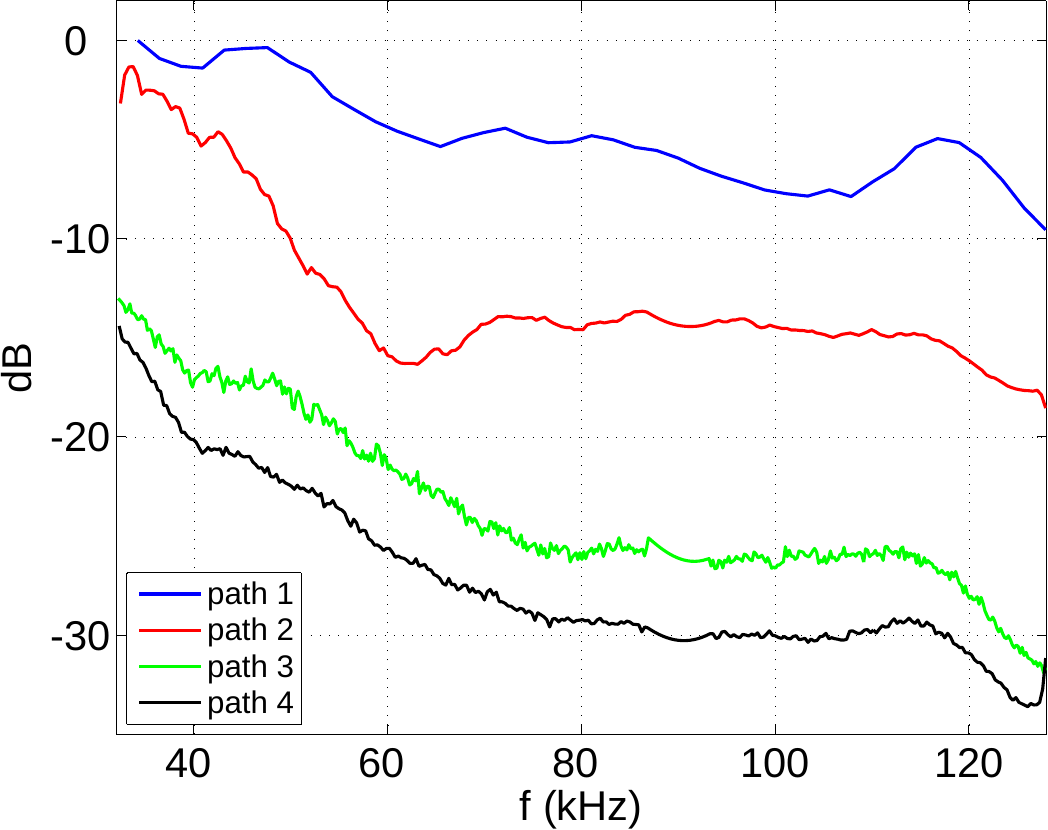}
    \caption{\label{fig:H_canal_1}}
  \end{subfigure} \\
	\vspace{4mm}
  \begin{subfigure}[t]{1\linewidth}
    \includegraphics[width=1\textwidth,height=0.81\textwidth]{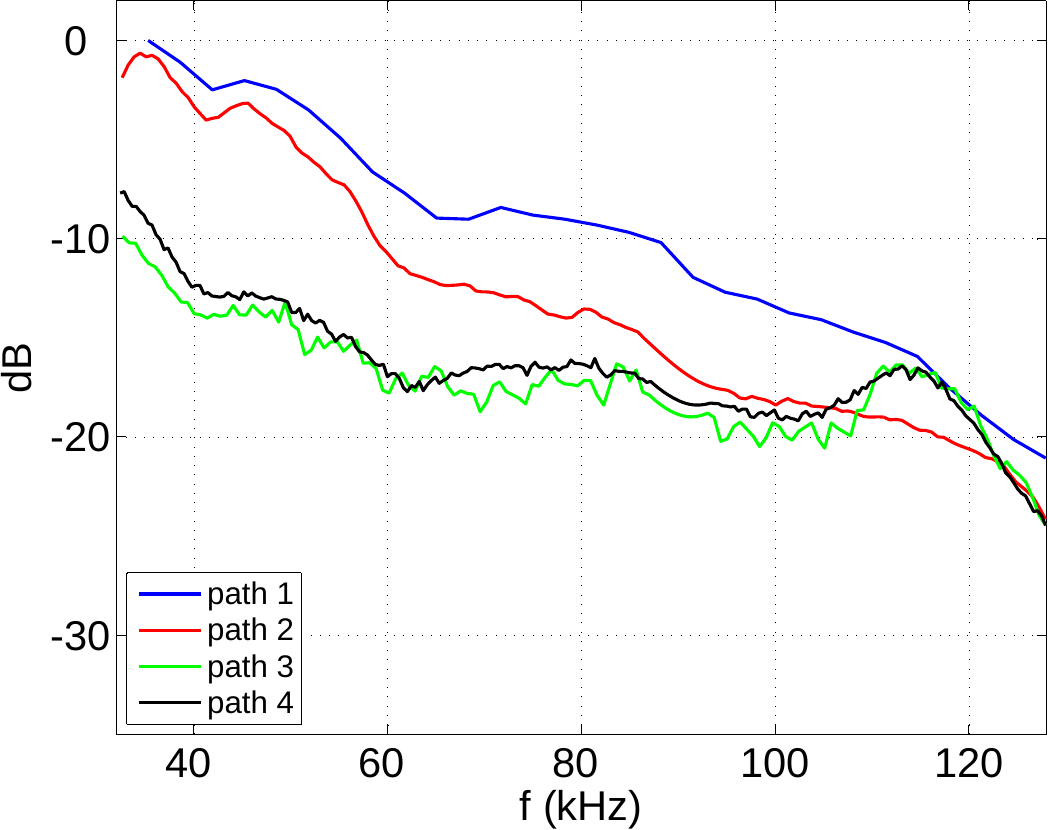}
    \caption{\label{fig:H_canal_8}}
  \end{subfigure}
\caption{Average frequency response $<|\tilde{H}_p(t,f)|>_t$ of the first four paths in (a) channel 1 and (b) channel 8.} 
\label{Fig_Rayos_2x2_H}
\end{figure}

\begin{figure}[t]
  \centering
  \begin{subfigure}[t]{1\linewidth}
    \includegraphics[width=1\textwidth]{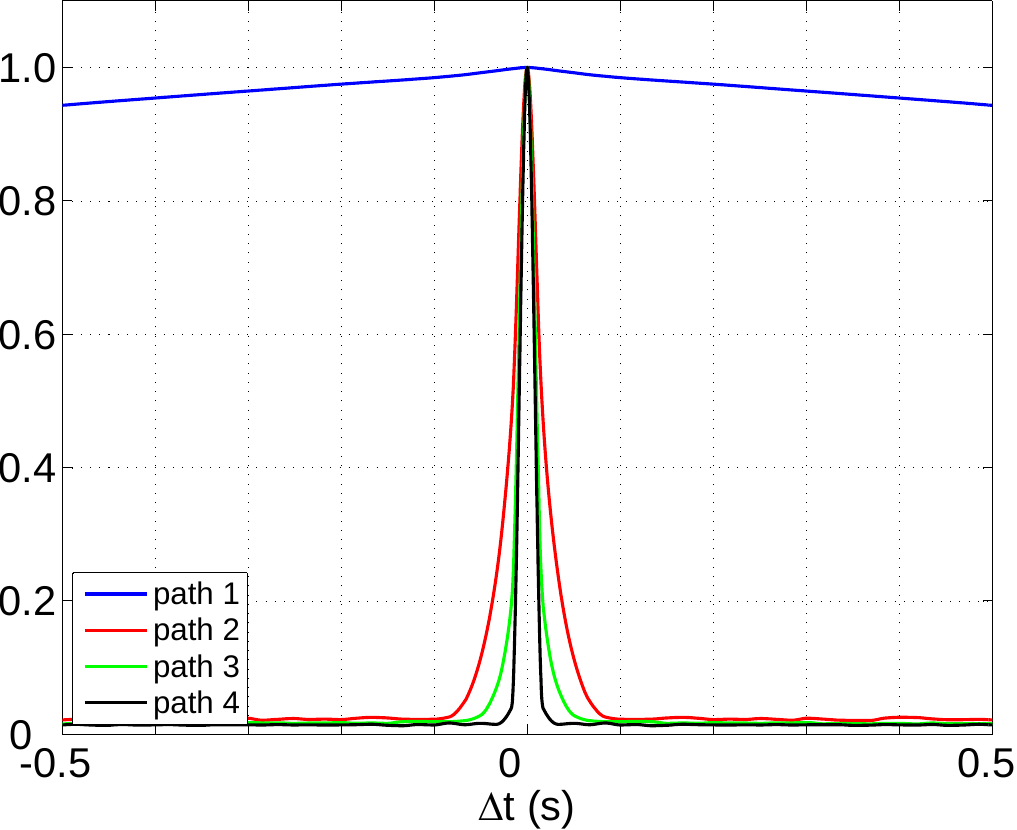}
    \caption{\label{Fig_Rayos_2x2_Autocorr_t_a}}
  \end{subfigure} \\
	\vspace{4mm}
  \begin{subfigure}[t]{1\linewidth}
    \includegraphics[width=1\textwidth]{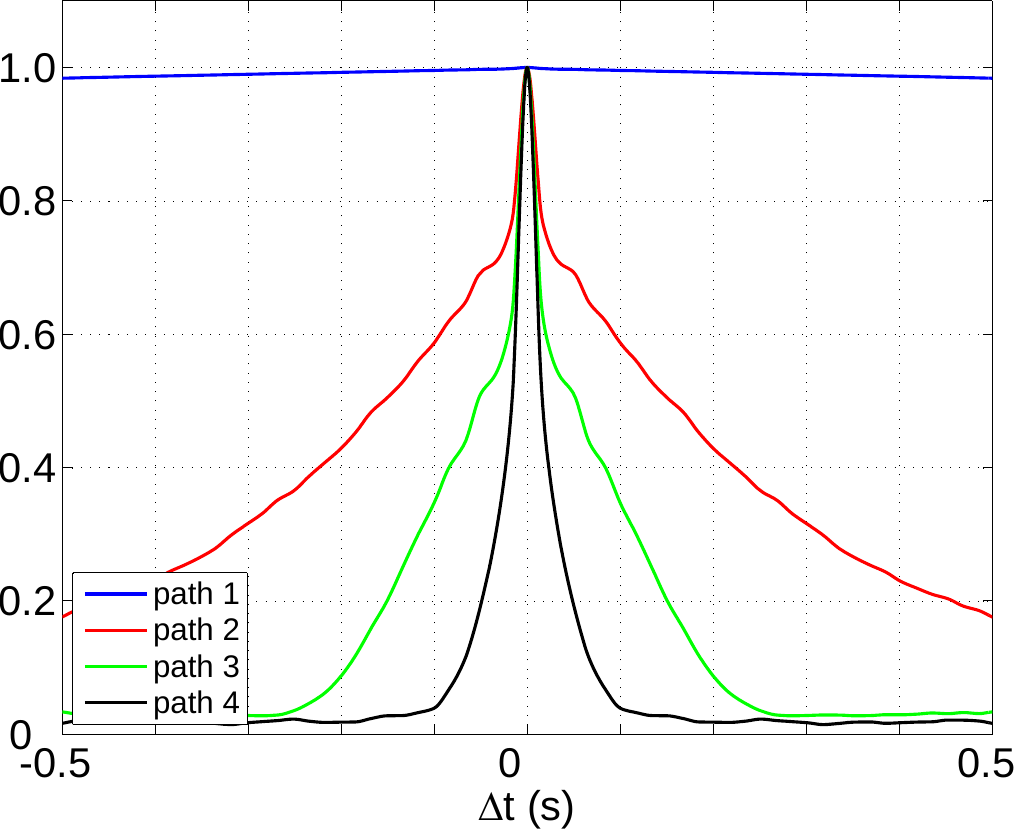}
    \caption{\label{Fig_Rayos_2x2_Autocorr_t_b}}
  \end{subfigure}
\caption{Absolute value of the time autocorrelation $|\Phi^{(t)}(\Delta t)|$ of the first 4 paths in (a) channel 1 and (b) channel 13.} 
\label{Fig_Rayos_2x2_Autocorr_t}
\end{figure}

\begin{figure}[t]
  \centering
  \begin{subfigure}[t]{1\linewidth}
    \includegraphics[width=1\textwidth]{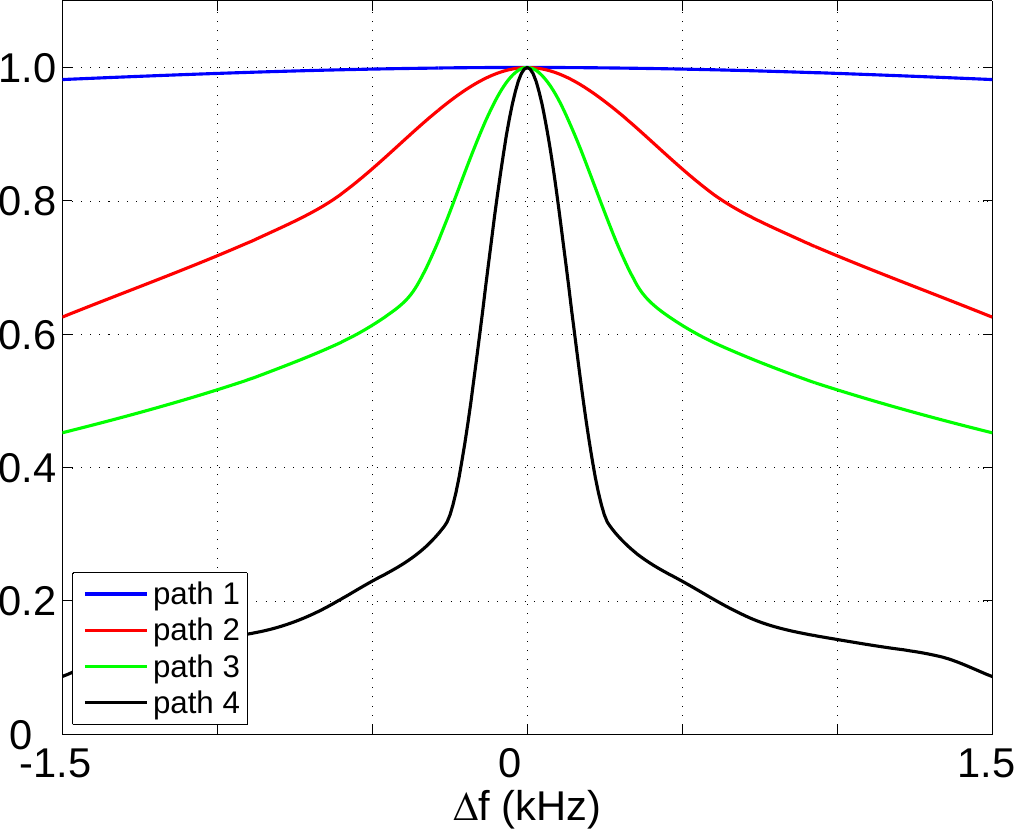}
    \caption{\label{Fig_Rayos_1x1_Autocorr_f_canal_3_crop_a}}
  \end{subfigure} \\
	\vspace{4mm}
  \begin{subfigure}[t]{1\linewidth}
    \includegraphics[width=1\textwidth]{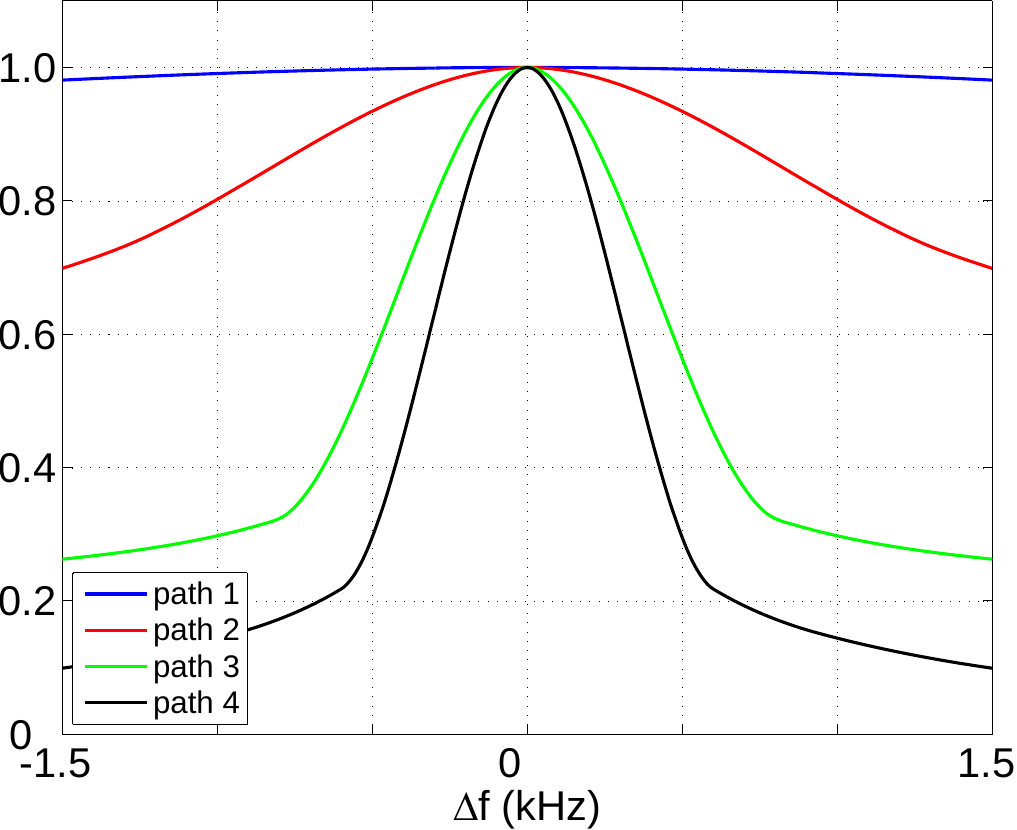}
    \caption{\label{Fig_Rayos_1x1_Autocorr_f_canal_10_crop_b}}
  \end{subfigure}
\caption{Absolute value of the frequency autocorrelation $|\Phi^{(f)}(\Delta f)|$ of the first 4 paths in (a) channel 3 and (b) channel 10.} 
\label{Fig_Rayos_2x2_Autocorr_f}
\end{figure}

\begin{figure*}
\centering
\includegraphics[width=\textwidth]{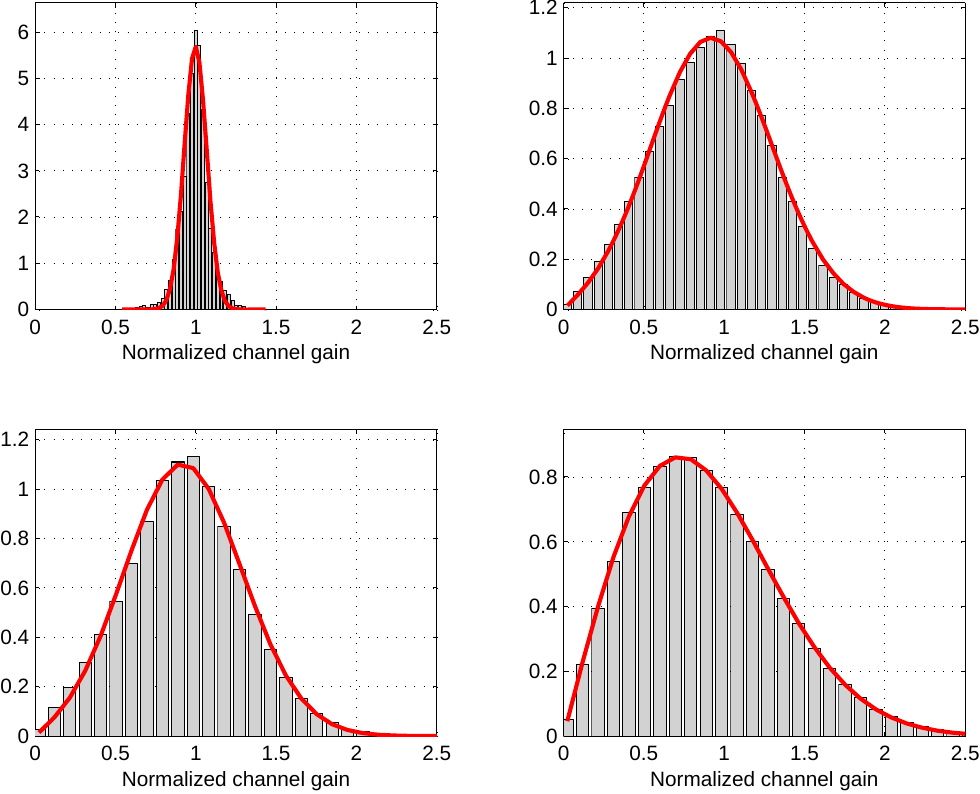}
\centering
\caption{Histogram of the normalized gains and fitted Rician pdf for the first 4 paths in channel 13.} 
\label{Fig_Rayos_2x2_HISTOGRAMA_C12}
\end{figure*}

\begin{table}       \caption{Global channel parameters}                                                                                                                    
\centering                                              
\begin{tabular}{|c|c|c|c|c|c|}                          
\hline                                                  
Channel & $t_c$ (ms) & $b_c$(kHz) & $\sigma_{\tau}$(ms) & K & $\epsilon$(\%) \\
\hline                                                  
1 & 52.4 & 1.0 & 2.1 & 2.5 & 5.2 \\                     
\hline                                                  
2 & 113.0 & 1.1 & 1.3 & 6.4 & 5.9 \\                    
\hline                                                  
3 & 36.4 & 2.0 & 2.8 & 0.9 & 0.8 \\                     
\hline                                                  
4 & 48.8 & 2.0 & 1.4 & 2.8 & 1.2 \\                     
\hline                                                  
5 & 58.2 & 2.6 & 1.6 & 1.9 & 0.4 \\                     
\hline                                                  
6 & 69.0 & 3.5 & 2.8 & 1.5 & 0.7 \\                     
\hline                                                  
7 & 45.6 & 4.0 & 2.8 & 0.8 & 0.5 \\                     
\hline                                                  
8 & 31.0 & 4.2 & 3.5 & 1.0 & 0.5 \\                     
\hline                                                  
9 & 88.2 & 4.8 & 1.8 & 2.3 & 1.7 \\                     
\hline                                                  
10 & 36.6 & 4.9 & 3.1 & 0.9 & 0.8 \\                    
\hline                                                  
11 & 68.8 & 5.1 & 3.4 & 1.4 & 0.9 \\                    
\hline                                                  
12 & 82.0 & 5.3 & 2.1 & 1.6 & 2.1 \\                    
\hline                                                  
13 & 69.5 & 7.8 & 2.4 & 1.6 & 2.3 \\                    
\hline                                                  
\end{tabular}                                                                       
\label{table:Global channel parameters}                              
\end{table} 

\subsection{Multipath channel characterization}
\label{Multipath channel characterization}
Results presented in \ref{Global channel characterization} are related to the global channel responses. The high frequencies used in the measurements allow us to go one step deeper and gain insight into the multipath nature of the channel by analyzing each path separately, as was done in \cite{Stojanovic2013} where an early assumption was made consistent in all the paths having an impulse response of the same shape. In our work, the results of the measurements evidence a considerable difference among the impulse responses and a clear non-flat frequency response of each path. 

We assume that the time-variant impulse response can be written as a sum of the impulse responses of every path as
\begin{equation}
\label{Eq:Modelo_multipath_t}
\tilde{h}(t,\tau)=\sum_{p=1}^{M}\tilde{h}_p(t,\tau-\tau_p),
\end{equation}
where $M$ is the number of significant paths, $\tilde{h}_p(t,\tau)$ is the time-variant impulse response of the p-\textit{th} path and $\tau_p$ is the delay where the p-\textit{th} path begins.
Taking the Fourier transform of (\ref{Eq:Modelo_multipath_t}) we get
\begin{equation}
\label{Eq:Modelo_multipath_f}
\tilde{H}(t,f)=\sum_{p=1}^{M}\tilde{H}_p(t,f)e^{-j2\pi f \tau_p},
\end{equation}
where $\tilde{H}_p(t,f)$ stands for the time-variant frequency response of the p-\textit{th} path. In our measurements the 4 first paths are clearly distinguishable while the rest have a negligible energy so they are not taken into account in our analysis.

The autocorrelations defined in (\ref{Eq:Autocorr_t_1}) and (\ref{Eq:Autocorr_f_1}) as well as the related coherence time and coherence bandwidth parameters will be applied to individually analyze each path inside a channel. We will also explore the statistical fluctuation of the gain of each path, as was done in section \ref{Global channel characterization}. 

In Fig. \ref{Fig_Rayos_2x2_H} we plot the 
absolute value of the average frequency response of the first four paths in channels 1 and 8. In particular in each subfigure we represent $20\log_{10}\left(<|\tilde{H}_p(t,f)|>_t\right)$, $p = 1, 2, 3, 4$. The plots are normalized to the maximum amplitude of the frequency response of the first (direct) path. Observe the great disparity of levels when changing from one channel to the other regarding both the absolute value of the frequency response of any of the paths and the level difference among the paths.

Fig. \ref{Fig_Rayos_2x2_Autocorr_t_a} and \ref{Fig_Rayos_2x2_Autocorr_t_b} show the absolute value of the time autocorrelation $|\Phi^{(t)}(\Delta t)|$ of the first 4 paths in channels 1 and 13 respectively. See that in general terms the functions decay slower for shorter paths. Indeed, the autocorrelation of first path decays at an extremely low rate so that it appears to be constant for the time span used in the figures, what turns into a coherence time of several seconds. This information is summarized in table \ref{table:Coherence time of the first 4 paths} where the coherence time of the first 4 paths for all the channels is shown under the labels $t_{c1}$, $t_{c2}$, $t_{c3}$ and $t_{c4}$. The fact that the direct path does not suffer surface or bottom reflection yields a remarkably high time coherence as opposed to the rest of the paths where the coherence time drastically decreases.

\begin{table}           
\caption{Coherence time of the first 4 paths.}                                                 
\centering                                                           
\begin{tabular}{|c|c|c|c|c|}                                         
\hline                                                               
Channel & $t_{c1}$(s) & $t_{c2}(ms)$ & $t_{c3}(ms)$ & $t_{c4}(ms)$ \\
\hline                                                               
1 & 1.1 & 5.4 & 4.0 & 3.4 \\                                         
\hline                                                               
2 & $>$2& 8.4 & 4.0 & 4.2 \\                                           
\hline                                                               
3 & $>$2& 5.8 & 5.6 & 4.6 \\                                           
\hline                                                               
4 & $>$2& 8.0 & 3.6 & 3.8 \\                                           
\hline                                                               
5 & $>$2& 9.2 & 3.6 & 3.8 \\                                           
\hline                                                               
6 & $>$2& 6.8 & 4.2 & 4.0 \\                                           
\hline                                                               
7 & $>$2& 13.5 & 5.4 & 6.2 \\                                          
\hline                                                               
8 & 1.2 & 5.0 & 4.2 & 4.2 \\                                         
\hline                                                               
9 & $>$2& 8.8 & 4.4 & 4.2 \\                                           
\hline                                                               
10 & 1.6 & 8.8 & 4.6 & 4.2 \\                                        
\hline                                                               
11 & $>$2& 6.0 & 3.8 & 3.8 \\                                          
\hline                                                               
12 & $>$2& 8.8 & 5.6 & 4.6 \\                                          
\hline                                                               
13 & $>$2& 8.2 & 6.2 & 5.4 \\                                          
\hline                                                               
\end{tabular}                                                        
\label{table:Coherence time of the first 4 paths}                                           
\end{table}

In Fig. \ref{Fig_Rayos_1x1_Autocorr_f_canal_3_crop_a} and \ref{Fig_Rayos_1x1_Autocorr_f_canal_10_crop_b} the corresponding absolute value of frequency autocorrelation $|\Phi^{(f)}(\Delta f)|$ for the first 4 paths for channels 3 and 10 is depicted. Again, the functions decay faster as we move from path 1 to path 4 in all the instances. In table \ref{table:Coherence bandwidth of the first 4 paths} the values of the coherence bandwidth for each path $b_{c1}$, $b_{c2}$, $b_{c3}$ and $b_{c4}$ and for all the channels is shown. Apart from the first paths which have a high coherence bandwidth of several kHz, the rest of the paths show a low coherence bandwidth and hence a pronounced frequency selectivity, as opposed to the narrowband results presented in \cite{Stojanovic2013} where the frequency response of the different paths is assumed to be flat.

\begin{table}                                                           
\caption{Coherence bandwidth of the first 4 paths.}                                                                                                          
\centering                                                              
\begin{tabular}{|c|c|c|c|c|}                                            
\hline                                                                  
Channel & $b_{c1}$(kHz) & $b_{c2}$(kHz) & $b_{c3}$(Hz) & $b_{c4}$(Hz) \\
\hline                                                                  
1 & 5.1 & 0.2 & 47.9 & 66.9 \\                                          
\hline                                                                  
2 & 7.3 & 0.3 & 76.5 & 274.6 \\                                         
\hline                                                                  
3 & 5.0 & 0.4 & 152.9 & 67.0 \\                                         
\hline                                                                  
4 & 6.4 & 0.4 & 57.4 & 105.1 \\                                         
\hline                                                                  
5 & 5.6 & 0.4 & 76.5 & 124.0 \\                                         
\hline                                                                  
6 & 6.4 & 0.4 & 143.1 & 95.5 \\                                         
\hline                                                                  
7 & 6.0 & 1.4 & 418.2 & 825.1 \\                                        
\hline                                                                  
8 & 4.0 & 0.3 & 171.5 & 95.5 \\                                         
\hline                                                                  
9 & 7.0 & 0.5 & 124.2 & 209.1 \\                                        
\hline                                                                  
10 & 4.9 & 0.6 & 209.3 & 143.0 \\                                       
\hline                                                                  
11 & 7.6 & 0.5 & 190.5 & 143.0 \\                                       
\hline                                                                  
12 & 8.4 & 0.5 & 257.1 & 293.7 \\                                       
\hline                                                                  
13 & 7.6 & 0.8 & 360.7 & 434.0 \\                                       
\hline                                                                  
\end{tabular}                                                                                                         
\label{table:Coherence bandwidth of the first 4 paths}                                              
\end{table}

Finally, the short-term channel gain statistic behavior for every path has been explored in the same manner as was done for the global channel in section \ref{Global channel characterization}. Fig.\ref{Fig_Rayos_2x2_HISTOGRAMA_C12} shows the histogram of the normalized gains of the first four paths in channel 13. Superimposed are the fitted Rician pdf. As could be expected, the histogram of the first path is much narrower than the rest, due to the lack of surface/bottom reflections that might lead to significant gain fluctuation. Following this argument, see how the histograms of paths 2, 3 and 4 show a growing dispersion. Table \ref{table:Rician K factor of the first 4 paths} shows the K factor of the Rician pdf that best fits the corresponding histogram in MSE sense as shown in (\ref{Eq:MSE}). A high K factor is obtained for the first paths in all the channels, in accordance to the behavior previously explained while a very low value of K is obtained for the rest of the paths which means a Rayleigh behavior. The relative fit error $\epsilon$ (not shown in the table for simplicity) is around 15\% for the first paths and less than 0.5\% for the rest.

\begin{table}    
\caption{Rician K factor of the first 4 paths.}                         
\centering                                
\begin{tabular}{|c|c|c|c|c|}              
\hline                                    
Channel & $K_1$ & $K_2$ & $K_3$ & $K_4$ \\
\hline                                    
1 & 11.2 & 0.0 & 0.2 & 0.2 \\             
\hline                                    
2 & 237.5 & 0.5 & 0.5 & 0.2 \\            
\hline                                    
3 & 46.3 & 0.6 & 1.6 & 0.3 \\             
\hline                                    
4 & 182.8 & 0.4 & 0.2 & 0.4 \\            
\hline                                    
5 & 158.3 & 0.5 & 0.3 & 0.5 \\            
\hline                                    
6 & 82.9 & 0.4 & 1.0 & 0.2 \\             
\hline                                    
7 & 121.4 & 2.0 & 5.6 & 0.5 \\            
\hline                                    
8 & 20.3 & 0.5 & 1.2 & 0.2 \\             
\hline                                    
9 & 152.1 & 0.9 & 1.1 & 0.3 \\            
\hline                                    
10 & 20.4 & 0.2 & 0.8 & 0.2 \\            
\hline                                    
11 & 122.6 & 1.1 & 1.3 & 0.2 \\           
\hline                                    
12 & 116.9 & 1.7 & 2.0 & 0.0 \\           
\hline                                    
13 & 100.4 & 2.3 & 2.5 & 0.3 \\           
\hline                                    
\end{tabular}                             
\label{table:Rician K factor of the first 4 paths}                
\end{table}

\section{Conclusions}
\label{Conclusions}
In this work we have presented the results of a campaign of wideband measurements conveyed at the ultrasonic frequencies ranging from 32 to 128 kHz. Ad-hoc probe signals have been designed for this purpose and their use in the estimation of the time-variant channel impulse and frequency responses has been described in detail. The wideband nature of the measurements have allowed the characterization of the channel not only globally but also for the main individual paths separately with an accuracy that is far beyond what is found in previous published works. This characterization has involved functions like the time and frequency autocorrelation, Doppler spectrum, delay profile and parameters like the coherence time and bandwidth, Doppler spread and delay spread. The random fluctuation of the channel gain has also been addressed resulting in a characterization of the shallow water channel not previously reported in the literature.

\section*{Acknowledgment}
This work has been partially supported by FEDER under the projects TEC2014-57901-R and UMA18-FEDERJA-085.

\appendix[Derivation of expression (\ref{Eq:Autocorr_Salida})]

The autocorrelation $\Phi^{(t)}_y(\Delta t)$ in (\ref{Eq:Autocorr_Salida}) is given by
\begin{equation}
\begin{split}
&\Phi^{(t)}_y(\Delta t)=<y(t)y(t-\Delta t)>_t=\\
\!\!&<\!\!\!\sum_{k=k_1}^{k_N} \!\! \Re\{\!H( t,k\delta_f)e^{j \theta_1(k)}\}\!\cdot \!  \!\!\! \sum_{k'=k_1}^{k_N}\!\!\! \Re\{\!H(\! t-\!\!\Delta t,k'\delta_f)e^{j\theta_2(k')} \}\!\!>_t,
\end{split} \label{Eq1}
\end{equation}

\vspace{-3mm}
where $\theta_1(k)$ and $\theta_2(k')$ are defined as
\begin{equation*}
\begin{split}
\theta_1(k) &\equiv 2\pi k \delta_f t+\Psi_k, \\
\theta_2(k') &\equiv 2\pi k' \delta_f (t-\Delta t)+\Psi_{k'}.
\end{split} 
\end{equation*}

Taking into account that 
\begin{equation*}
\Re\{z_1\}\cdot \Re\{z_2\}=\tfrac{1}{2}\Re\{z_1 z_2\}+\tfrac{1}{2}\Re\{z_1 z_2^*\},
\end{equation*}
\vspace{-1mm}
 where $z_1$ and $z_2$ are complex numbers we can express the autocorrelation $\Phi^{(t)}_y(\Delta t)$ in (\ref{Eq1}) as
\begin{equation}\label{Eq:2}
\begin{split}
&\Phi^{(t)}_y(\Delta t)= \\
\!\!\!\!&\tfrac{1}{2}\sum_{k=k_1}^{k_N}\!\sum_{k'=k_1}^{k_N}\!\!\!\!<\Re\{ H(t,k\delta_f) H(t\!\!-\!\!\Delta t,k'\delta_f) e^{j(\theta_1(k)+\theta_2(k'))} \}\!\!>_t + \!\\
&\qquad \qquad \!\!<\Re\{H(t,k\delta_f) H^*(t\!\!-\!\!\Delta t,k'\delta_f) e^{j(\theta_1(k)-\theta_2(k'))} \}\!\!>_t 
\end{split} 
\end{equation}
\vspace{-1mm}
where 
\begin{equation} \label{Eq:Phis}
\begin{split}
\theta_1(k) + \theta_2(k') &= 2\pi (k+k') \delta_f t -2\pi k' \delta_f \Delta t +\Psi_k+\Psi_{k'}, \\
\theta_1(k) - \theta_2(k') &= 2\pi (k-k') \delta_f t +2\pi k' \delta_f \Delta t +\Psi_{k}-\Psi_{k'}.
\end{split} 
\end{equation}
\vspace{-1mm}
Let's assume that $H(t,f)$ varies with time much slower than any of the complex exponentials in (\ref{Eq:2}), which have the general form $e^{j2\pi m\delta_f t+\psi}$. This means that $H(t,f)$ can be considered as a constant in any period $1/(m\delta_f)$ and therefore, the time average of every addend in (\ref{Eq:2}) is approximately zero except for the case $m=0$, i.e. $k'=k$ in (\ref{Eq:Phis}). With this assumption we can simplify (\ref{Eq:2}) resulting in the expression shown in (\ref{Eq:Autocorr_Salida}), this is

\vspace{-3mm}
\begin{equation*}
\begin{split}
&\Phi^{(t)}_y(\Delta t)\approx \\
&\sum_{k=k_1}^{k_N}\tfrac{1}{2}\Re\{<H(t,k\delta_f) H^*(t-\Delta t,k\delta_f)>_t e^{j(2\pi k\delta_f\Delta t )}\} =\\
&\sum_{k=k_1}^{k_N}\tfrac{1}{2}\Re\{\Phi^{(t)}_H (\Delta t,k\delta_f)e^{j(2\pi k\delta_f\Delta t )}\}.
\end{split}
\end{equation*}

\bibliographystyle{ieeetr}
\bibliography{Underwater}

\begin{thebibliography}{10}

\bibitem{Walree2010}
H.~S. Paul Van~Walree, Trond~Jenserud, ``Characterization of overspread
  acoustic communication channels,'' in {\em 10th European Conference on
  Underwater Acoustics}, pp.~952--958, Jul 2010.

\bibitem{StojanovicCommMag2009}
M.~Stojanovic and J.~Preisig, ``Underwater acoustic communication channels:
  Propagation models and statistical characterization,'' {\em IEEE
  Communications Magazine}, vol.~47, pp.~84--89, January 2009.

\bibitem{Yang2006}
W.-B. Yang and T.~C. Yang, ``High-frequency channel characterization for m-ary
  frequency-shift-keying underwater acoustic communications,'' {\em The Journal
  of the Acoustical Society of America}, vol.~120, no.~5, pp.~2615--2626, 2006.

\bibitem{Kim2013}
S.~{Kim}, S.~{Byun}, S.~{Kim}, , and and, ``Underwater acoustic channel
  characterization at 6khz and 12khz in a shallow water near jeju island,'' in
  {\em 2013 OCEANS - San Diego}, pp.~1--4, Sep. 2013.

\bibitem{Kulhandjian2014}
H.~Kulhandjian and T.~Melodia, ``Modeling underwater acoustic channels in
  short-range shallow water environments,'' in {\em Proceedings of the
  International Conference on Underwater Networks \& Systems}, WUWNET '14, (New
  York, NY, USA), pp.~26:1--26:5, ACM, 2014.

\bibitem{Stojanovic2013}
P.~Qarabaqi and M.~Stojanovic, ``Statistical characterization and
  computationally efficient modeling of a class of underwater acoustic
  communication channels,'' {\em IEEE Journal of Oceanic Engineering}, vol.~38,
  pp.~701--717, Oct 2013.

\bibitem{Kim2015}
J.~Kim, I.-S. Koh, and Y.~Lee, ``Short-term fading model for signals reflected
  by ocean surfaces in underwater acoustic communication,'' {\em IET
  Communications}, vol.~9, pp.~1147--1153(6), June 2015.

\bibitem{Walree2013}
P.~A. van Walree and R.~Otnes, ``Ultrawideband underwater acoustic
  communication channels,'' {\em IEEE Journal of Oceanic Engineering}, vol.~38,
  pp.~678--688, Oct 2013.

\bibitem{Walree2013bis}
P.~A. van Walree, ``Propagation and scattering effects in underwater acoustic
  communication channels,'' {\em IEEE Journal of Oceanic Engineering}, vol.~38,
  pp.~614--631, Oct 2013.

\bibitem{Chitre2007}
M.~Chitre, ``A high-frequency warm shallow water acoustic communications
  channel model and measurements,'' {\em The Journal of the Acoustical Society
  of America}, vol.~122, no.~5, pp.~2580--2586, 2007.

\bibitem{Zhang2010}
J.~Zhang, J.~Cross, and Y.~R. Zheng, ``Statistical channel modeling of wireless
  shallow water acoustic communications from experiment data,'' in {\em 2010 -
  MILCOM 2010 MILITARY COMMUNICATIONS CONFERENCE}, pp.~2412--2416, Oct 2010.

\bibitem{Ochi2008}
H.~{Ochi}, Y.~{Watanabe}, T.~{Shimura}, and T.~{Hattori}, ``Experimental
  results of short range wideband acoustic communication using qpsk and 8psk,''
  in {\em OCEANS 2008 - MTS/IEEE Kobe Techno-Ocean}, pp.~1--5, April 2008.

\bibitem{Hajenko2010}
T.~J. {Hajenko} and C.~R. {Benson}, ``The high frequency underwater acoustic
  channel,'' in {\em OCEANS'10 IEEE SYDNEY}, pp.~1--3, May 2010.

\bibitem{Canete2016}
F.~J. Cañete, J.~López-Fernández, C.~García-Corrales, A.~Sánchez,
  E.~Robles, F.~J. Rodrigo, and J.~F. Paris, ``Measurement and modeling of
  narrowband channels for ultrasonic underwater communications,'' {\em
  Sensors}, vol.~16, no.~2, 2016.

\bibitem{Jurgen2013}
A.~F. Hans-Jurgen~Zepernick, {\em Pseudo Random Signal Processing: Theory and
  Application}.
\newblock Wiley-Blackwell, 2013.

\bibitem{Vaidyanathan1993}
P.~P. Vaidyanathan, {\em Multirate Systems and Filter Banks}.
\newblock Upper Saddle River, NJ, USA: Prentice-Hall, Inc., 1993.

\bibitem{Bello1963}
P.~{Bello}, ``Characterization of randomly time-variant linear channels,'' {\em
  IEEE Transactions on Communications Systems}, vol.~11, pp.~360--393, December
  1963.

\bibitem{simon2005}
M.~K. Simon and M.-S. Alouini, {\em Digital communication over fading
  channels}, vol.~95.
\newblock John Wiley \& Sons, 2005.

\bibitem{Cirrone2004}
G.~A.~P. {Cirrone}, S.~{Donadio}, S.~{Guatelli}, A.~{Mantero}, B.~{Mascialino},
  S.~{Parlati}, M.~G. {Pia}, A.~{Pfeiffer}, A.~{Ribon}, and P.~{Viarengo}, ``A
  goodness-of-fit statistical toolkit,'' {\em IEEE Transactions on Nuclear
  Science}, vol.~51, pp.~2056--2063, Oct 2004.

\bibitem{Lin1991}
J.~{Lin}, ``Divergence measures based on the shannon entropy,'' {\em IEEE
  Transactions on Information Theory}, vol.~37, pp.~145--151, Jan 1991.

\end{thebibliography}

\end{document}